\documentclass[twocolumn]{aastex631}
\graphicspath{{./}{figures/}}
\usepackage[]{CJK}

\begin{document}
\title{The frequency ratio and time delay of solar radio emissions with fundamental  and harmonic components}
\author[0000-0002-1810-6706]{Xingyao Chen (\begin{CJK}{UTF8}{gbsn}陈星瑶\end{CJK})}
\affiliation{School of Physics \& Astronomy, University of Glasgow, Glasgow, G12 8QQ, UK}
\author[0000-0002-8078-0902]{Eduard P. Kontar}
\affil{School of Physics \& Astronomy, University of Glasgow, Glasgow, G12 8QQ, UK}
\author[0000-0003-1967-5078]{Daniel L. Clarkson}
\affil{School of Physics \& Astronomy, University of Glasgow, Glasgow, G12 8QQ, UK}
\author[0000-0002-4389-5540]{Nicolina Chrysaphi}
\affiliation{LESIA, Observatoire de Paris, Universit\'{e} PSL, CNRS, Sorbonne Universit\'{e}, Universit\'{e} de Paris, 5 place Jules Janssen, 92195 Meudon, France}
\affiliation{School of Physics \& Astronomy, University of Glasgow, Glasgow, G12 8QQ, UK}

\correspondingauthor{Xingyao Chen}\email{Xingyao.Chen@glasgow.ac.uk}

\begin{abstract}
Solar radio bursts generated through the plasma emission mechanism produce radiation near the local plasma frequency (fundamental emission) and double the plasma frequency (harmonic). While the theoretical ratio of these two frequencies is close to 2, simultaneous observations give ratios ranging from 1.6 to 2, suggesting either a ratio different from 2, a delay of the fundamental emission, or both. To address this long-standing question, we conducted high frequency, high time resolution imaging spectroscopy of type III and type J bursts with fine structures for both the fundamental and harmonic components with LOFAR between 30 and 80 MHz. The short-lived and narrow frequency-band fine structures observed simultaneously at fundamental and harmonic frequencies give a frequency ratio of 1.66 and 1.73, similar to previous observations. However, frequency-time cross-correlations suggest a frequency ratio of 1.99 and 1.95 with a time delay between the F and H emissions of 1.00 and 1.67 s, respectively for each event. Hence, simultaneous frequency ratio measurements different from 2 are caused by the delay of the fundamental emission. Among the processes causing fundamental emission delays, anisotropic radio-wave scattering is dominant. Moreover, the levels of anisotropy and density fluctuations reproducing the delay of fundamental emissions are consistent with those required to simulate the source size and duration of fundamental emissions. Using these simulations we are able to, for the first time, provide quantitative estimates of the delay time of the fundamental emissions caused by radio-wave propagation effects at multiple frequencies, which can be used in future studies.
\end{abstract}

\section{Introduction}
\label{sec01}

Plasma emission is believed to be responsible for generating solar radio bursts at decimeter and longer wavelengths \citep[e.g.][]{1958SvA.....2..653G, 1985ARA&A..23..169D, 1987SoPh..111...89M}. 
Within the plasma emission theory, the instability generates Langmuir waves at the local plasma frequency $f_\mathrm{pe}$, where $f_\mathrm{pe}=\omega_\mathrm{pe}/2\pi=\sqrt{e^2 n (r)/\pi m_e}$ is the electron plasma frequency, $n(r)$ is the electron number density, and $e$ and $m_e$ are the electron charge and mass.
The coalescence of Langmuir waves and low frequency ion-sound waves may produce electromagnetic waves via nonlinear plasma processes, which is referred to as fundamental emission (hereafter, F). 
The coalescence of counter-propagating Langmuir waves may produce the radio emission at 2$f_\mathrm{pe}$, which is referred to as the second harmonic emission (hereafter, H). 

Several longstanding issues exist concerning the F and H emissions---one problem in particular is how to distinguish between them.
When a single burst is observed (and not an F-H pair), it is difficult to identify whether this burst is the result of F or H emissions, with the identification relying on the degree of their polarisation.
The polarisation degree of the F radio waves is predicted to be higher than the H \citep{1985ARA&A..23..169D}. 
\cite{1985srph.book..289S} reported that the polarisation degree of the F components are all larger than the H components from 714 F-H pairs of type III bursts, that being 0.35 (F) and 0.12 (H) on average. 
However, the polarisation degree varies even within one burst \citep{1980A&A....88..203D} or even within one fundamental component \citep{2017NatCo...8.1515K}.
When there are two components observed, some additional arguments can suggest F-H pairs: the F typically has higher intensities, shorter rise times, higher polarisations, and near half drift rates of the higher frequency component.

There are also multiple issues in debate---for example, F and H emission may be generated over different timescales  because they are produced through related, but different, processes. 
Moreover, it is unclear whether F and H emissions are produced at the same spatial location.
In this paper, we consider a remarkable problem related to their frequency ratio $R_\mathrm{H/F}$, i.e. the ratio of the H to F components observed at the same time. While the theory gives a harmonic frequency ratio very close to 2, the observations suggest a range from 1.6-2, averaging at 1.8 \citep{1954AuJPh...7..439W, 1974SoPh...39..451S}.

Multiple studies have previously indicated time delays of the F with respect to the H, in a range from 1 s up to 7 s, and H/F frequency ratios in a range of 1.74-1.94 \citep{1963ApJ...138..239H, 1974SoPh...39..451S, 1998SoPh..181..363R, 2015SoPh..290..181D, 2016ApJ...826..125K, 2018SoPh..293...26M}, yet there is no agreement on the mechanisms involved.
For example, an early explanation by \cite{1954AuJPh...7..439W} for the observed difference from the theoretical prediction being that the observed spectrum consists of the H with only the higher frequencies of the F band; \cite{1974SoPh...39..451S} suggested that the F emission had reduced escape efficiency; the model by \cite{1998SoPh..181..363R} stated that the F was reabsorbed above the local plasma frequency; \cite{2015SoPh..290..181D} suggested lower group velocities of the F inside a modeled magnetic loop.

There are few studies aimed at the analysis of the H/F frequency ratios, because firstly, the F emissions are not always observed along with their corresponding H emissions and are difficult to be clearly defined; secondly, the H/F frequency ratios are usually used as evidence for the H components, while they are close to 2 then the two emission branches may be regarded as F-H pairs; and thirdly, it is hard to quantitatively compare a F frequency with the corresponding H component in order to calculate the H/F frequency ratios.

From previous studies, the cause of the time delay of the F components has been mostly explained by their different electromagnetic wave group velocities and radio wave propagation effects in an inhomogeneous corona \citep{1993A&A...279..235I, 2018SoPh..293...26M}. However, there was no quantitative estimation of the H/F frequency ratios and delay times from radio wave propagation effects.
There seems to be no obvious dependencies of H/F frequency ratios on frequencies from previous observations, which suggests that the propagation effects may be dominant due to various characteristic parameters of the background density fluctuations in a turbulent corona.

In this paper, we study the H/F frequency ratios and delay times between the F and H emissions from type III and type J bursts with striae observed by the LOw Frequency ARray \citep[LOFAR;][]{2013A&A...556A...2V} with high temporal, spectral and spatial resolutions in a frequency range of 30-80 MHz.
J-type and/or U-type solar radio bursts is a variant of type III bursts and is believed to be generated from electron beams traveling along closed magnetic loops \citep{1958Natur.181...36M, 1970PASA....1..316L,1990A&A...229..216H,1992ApJ...391..380A, 1997A&AS..123..279A,2010AIPC.1206..433D,2012AdSpR..49.1607F,2017A&A...606A.141R}.
From a featureless radio burst with F and H components, the frequency ratio can be calculated from the frequencies at the maximum intensities of F and H components at each time. The time delay can be derived from the times at the peak intensities at $f$ and 2$f$. However, it does not seem possible to determine the H/F frequency ratio and delay time simultaneously.
We implement the correlation of type III and type J bursts that present striae in both F and H components, allowing clear determination of the frequency ratio. 
The fine structures are crucial to this method because striae in the F components have their counterparts in the H components and can be well correlated to determine the frequency ratio and delay time simultaneously.
The variants of type III bursts in a form of `J' or `U' in the dynamic spectrum can also give the frequency ratio and delay time simultaneously from their similar shapes but it is worth noting that they are different if measured at the turning point and starting frequency of type U or J bursts. The isolines with different intensity levels of the U or J structure would also affect the estimations of frequency ratio and delay time.

\begin{figure*}[t!] 
\epsscale{1.}
\plotone{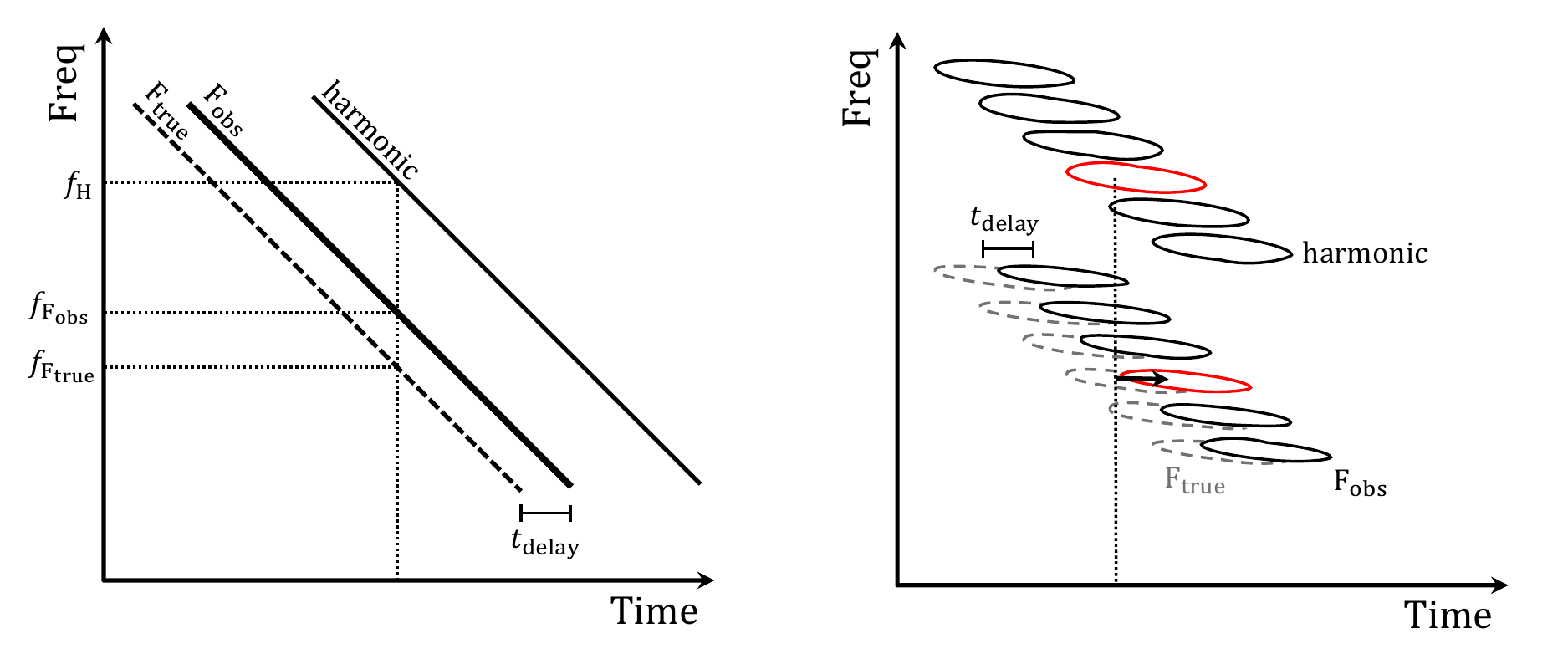}
\caption{A physical scenario of two dynamic spectra showing the time delay between the observed and intrinsic F components of the normal type III burst (left panel) and the type III burst with striae fine structures (right panel).}
\label{0cartoon}
\end{figure*}

From the dynamic spectra of type III bursts with striae, the observed H/F frequency ratios are less than the theoretical ratio, suggesting that the observed F branch is delayed.
We suggest that the time delay between F and H components can be explained by a combination of different group velocities and scattering effects, with the latter forming a larger contribution.
As the F component is emitted at the local plasma frequency, it undergoes a stronger scattering effect than the H component, which lengthens the propagation path of the F emission and leads to a delayed F band.
For the first time, we quantitatively estimate the delay times between the F and H components from ray-tracing simulations of radio wave propagation \citep{2019ApJ...884..122K}.
Based on the diagnosis by \cite{2020ApJ...905...43C} in which they estimated the turbulent coronal parameters by combined analysis of scattering simulations and imaging observations from one type IIIb radio burst, our estimations of delay times from both different group velocities and scattering propagation effects can be well matched with that from observations.

In Section \ref{sec2}, we present the theoretical H/F frequency ratio.
A schematic illustration of the delayed F component in the type III burst with striae is presented in Section \ref{sec3}. 
Section \ref{sec4} describes the characteristic parameters from type III and type J burst observations, including their observed H/F frequency ratios and delay times.
Section \ref{sec5} shows the results from ray-tracing simulations of radio wave propagation. 
Section \ref{sec6} is our summary.

\section{Emission near fundamental and harmonic frequencies}
\label{sec2}

The plasma emission mechanism is widely accepted for the generation of solar radio bursts at decimeter and longer wavelengths \cite[e.g.][]{1958SvA.....2..653G,1985ARA&A..23..169D, 1985srph.book.....M, 1996ASSL..204.....Z, 2014A&A...572A.111R}. 
The F emission can be generated through scattering of Langmuir waves by ions and/or their interaction with ion-sound waves. The H emission is generated by the coalescence of the Langmuir waves with back-scattered Langmuir waves, so emits at the summed frequencies of two Langmuir waves.

Conservation of energy and momentum (frequency and wavenumber) leads to the fundamental emission being close to the Langmuir wave frequency 
\[
\omega_\mathrm{L}\simeq \omega _\mathrm{pe}\left(1+\frac{3}{2}\frac{v_\mathrm{Te}^2}{v^2}\right)
\]
where $v_\mathrm{Te}$ is the electron thermal speed, and $v$ is the phase speed of the waves.
For the typical parameters of type III solar radio bursts, $v/v_\mathrm{Te}$ is about $10$ \citep[e.g.][]{2014A&A...572A.111R,2021NatAs...5..796R}, 
so the deviation from plasma frequency $\Delta \omega$ is as small as
\[
\Delta \omega/\omega _\mathrm{pe}\simeq 0.015
\]
Similarly, the harmonic frequency is close to the twice of the Langmuir wave frequency. Hence the ratio of harmonic to fundamental should be very close to 2, within $1-2$\% for the typical parameters in type III solar radio bursts.

\section{Schematic illustration}
\label{sec3}

The schematic illustration outlined in Figure \ref{0cartoon} clearly explains a delayed F component leads to the derived H/F frequency ratio to be less than 2 from observations of F-H pairs. 
The observed F component is delayed by a time $t_\mathrm{delay}$ compared to the intrinsic F component. The frequency of the observed F branch is larger than the intrinsic F branch at each time, so the H/F frequency ratio normally calculated between the observed H and F components will be less than the theoretical frequency ratio. The intrinsic H/F frequency ratio should be between the observed H component and the intrinsic F component instead of the observed F component. Furthermore, the radio source imaging of the H component should be coincident with an earlier F branch instead of the F emission at the same time.

We also show a scenario for a type III burst with striae in the right panel from Figure \ref{0cartoon}. 
With the striae, we can determine the frequency ratio and delay time simultaneously, which is not feasible for normal type III bursts without fine structures.
Ideally, striae in the F components are expected to have counterparts in the H components, yet the striae in the H are normally not as apparent, which may be the result of the weak intensity of the H components and the limited dynamic range of antennas. Nonetheless, the striae in our analysis are well observed and correlated with both F and H components from LOFAR observations.

\begin{figure*}[t!]
\gridline{\fig{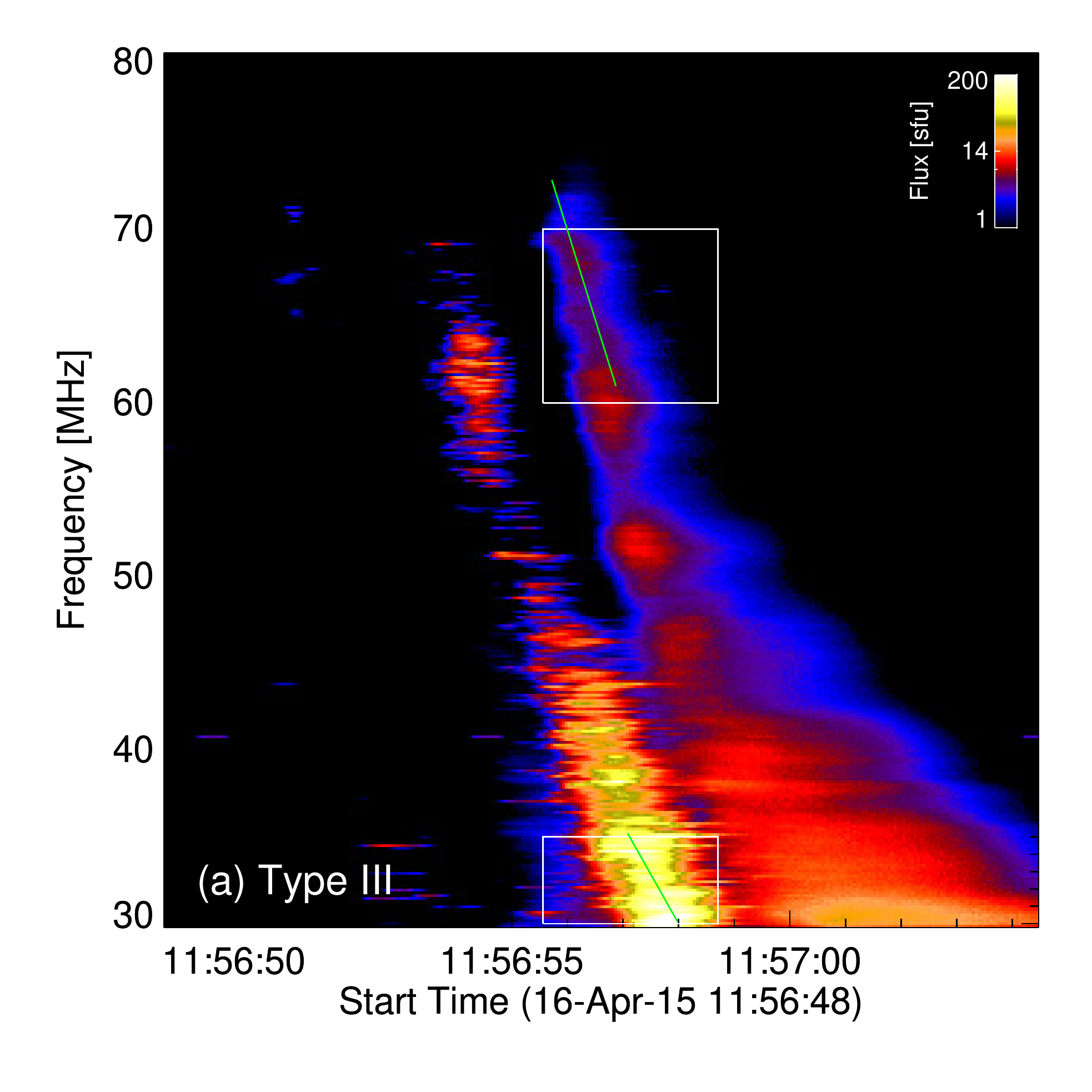}{0.45\textwidth}{ }
          \fig{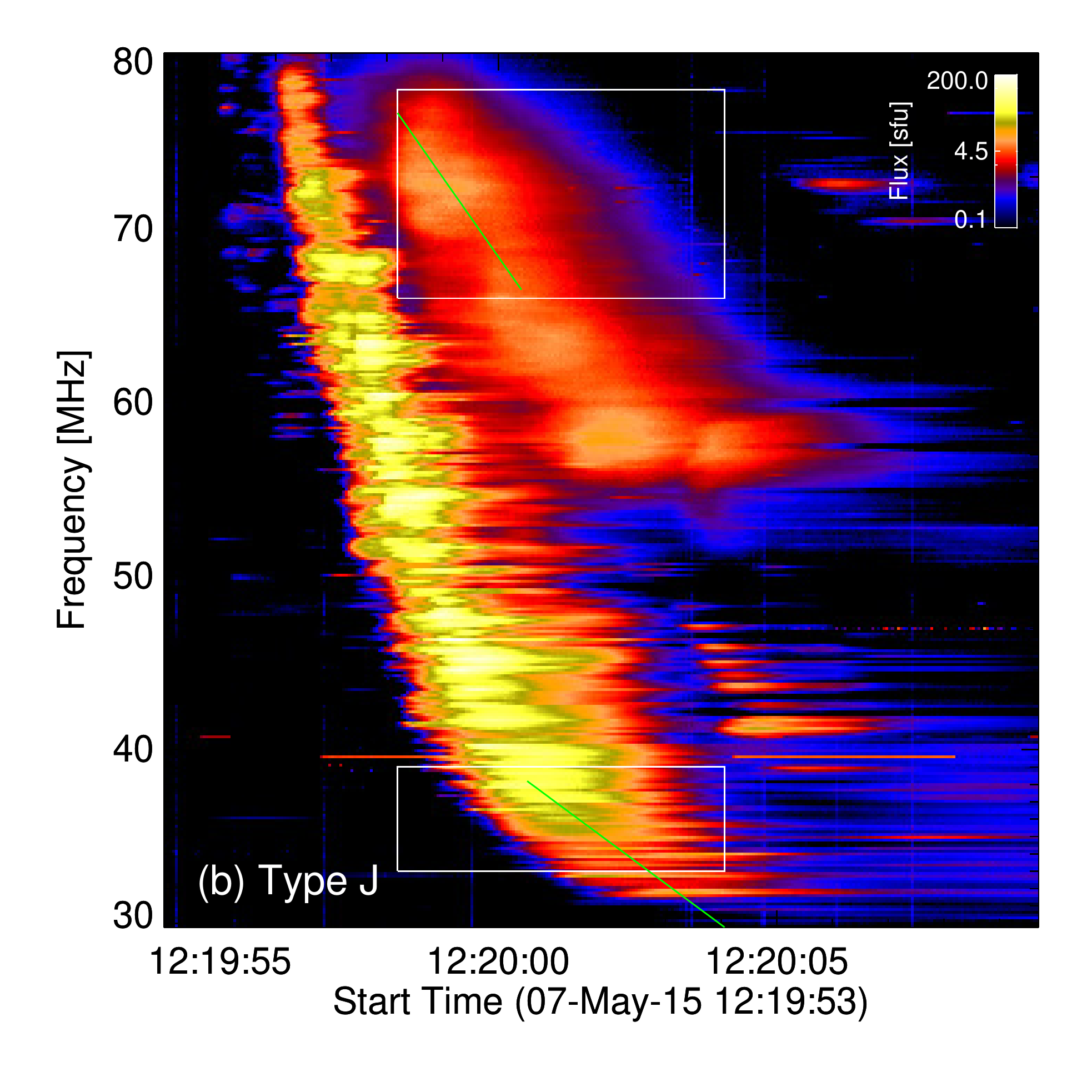}{0.45\textwidth}{ }}
\caption{The dynamic spectra of type III (a) and J (b) bursts with both F and H components observed by LOFAR. The same time range for both F and H components, but twice the frequency range for the H component, is selected to derive their drift rates, indicated by the solid rectangular boxes. The green lines are the best fit through all the positions of the fitted Gaussian peaks using a linear fitting function.}
\label{fig-spec}
\end{figure*}

\section{Observations}
\label{sec4}

The solar type III radio burst (left panel in Figure \ref{fig-spec}) at around 11:57:00 UT on 16-April-2015 and the type J burst (right panel in Figure \ref{fig-spec}) at 12:20:00 UT on 07-May-2015 are observed by the LOw Frequency ARray \citep[LOFAR;][]{2013A&A...556A...2V}.  
LOFAR is a large interferometric radio telescope with high spectroscopic and imaging capabilities, located primarily in the Netherlands with a number of international stations in other European countries. It was completed in 2012 by the Netherlands Institute for Radio Astronomy (ASTRON) and can observe with the Low Band Antenna (LBA) and the High Band Antenna (HBA), optimized for 30 - 80 MHz and 120 - 240 MHz, respectively. The type III and J bursts here are observed by the tied-array beam forming mode simultaneously with a maximum frequency resolution of $\sim$ 12.2 kHz and time resolution of $\sim$ 10 ms \citep{2017NatCo...8.1515K}.

We show our analysis of well observed type III and J bursts with both F and H components and striae, derive their frequency ratios and delay times, and compare with those estimated from radio wave propagation simulations.

\begin{figure*}
\gridline{\fig{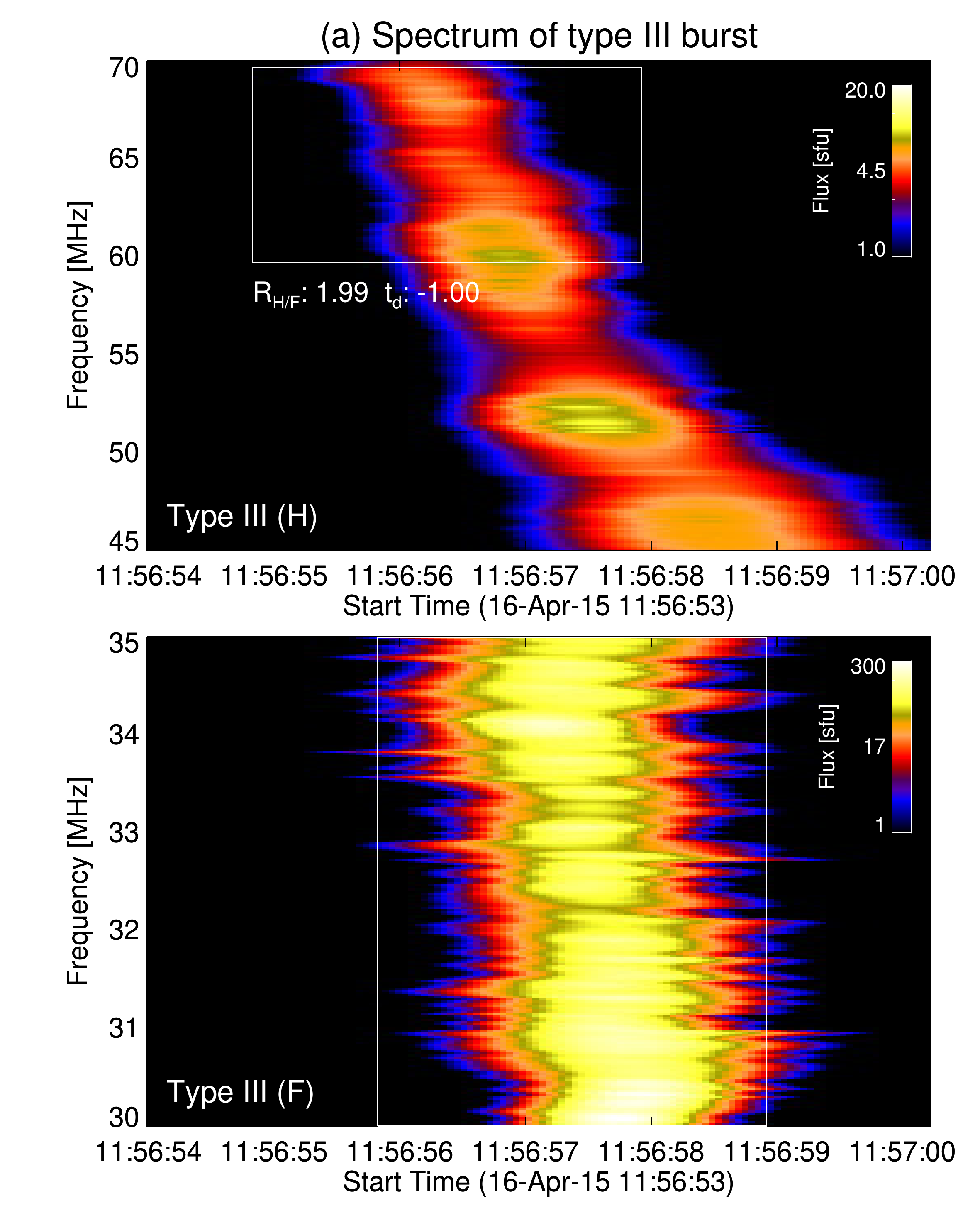}{0.45\textwidth}{ }
          \fig{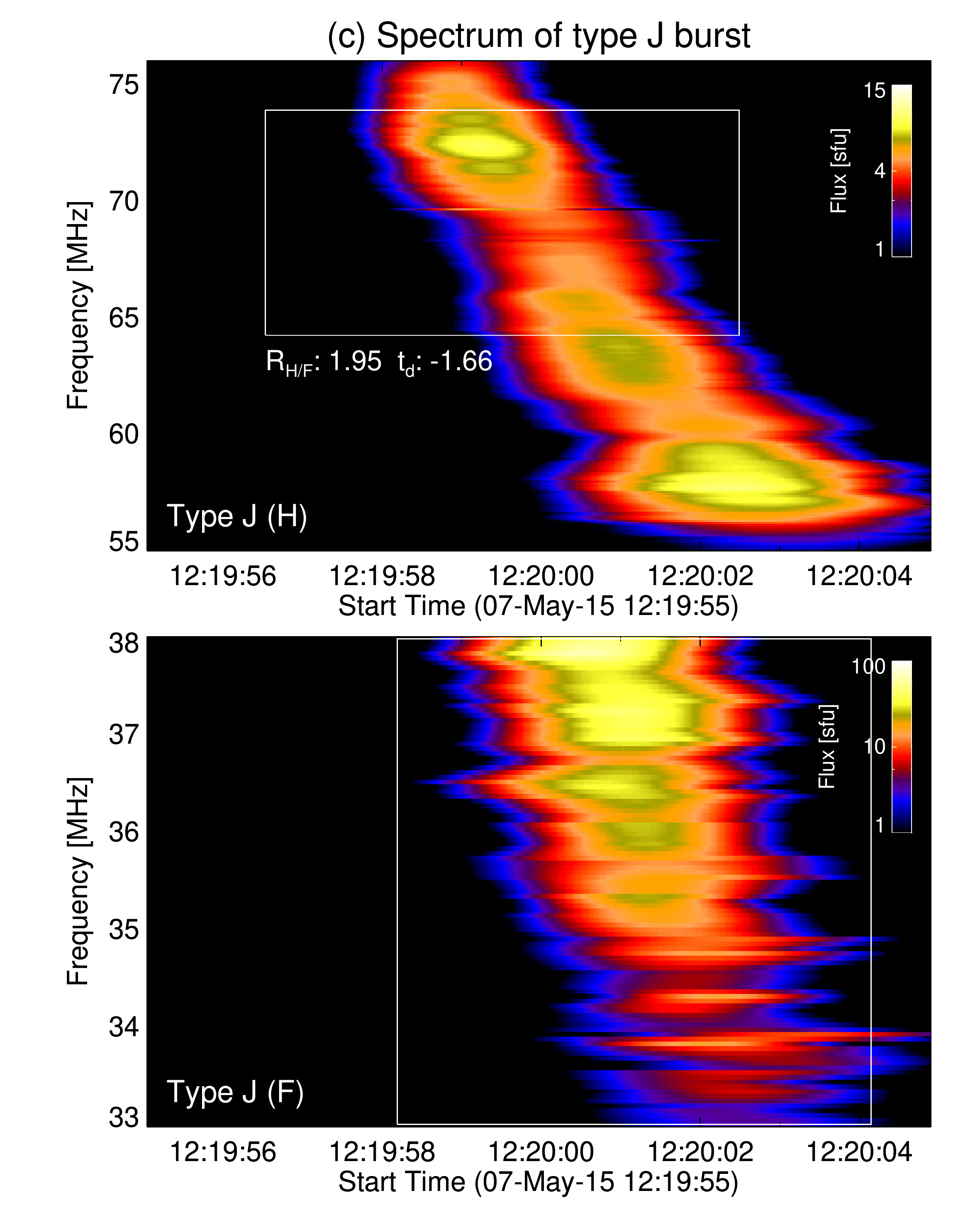}{0.45\textwidth}{ }}
\gridline{\fig{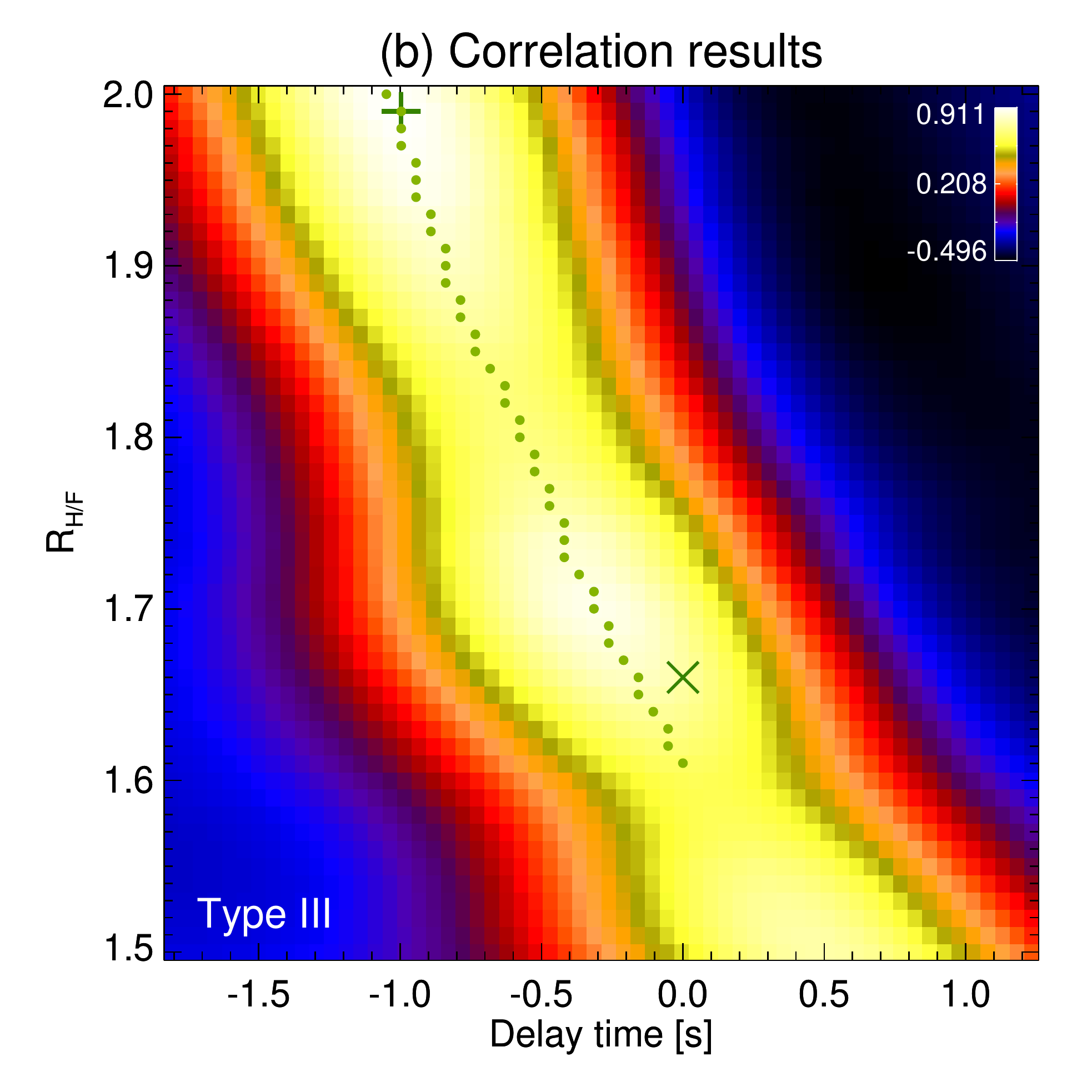}{0.41\textwidth}{ }
          \fig{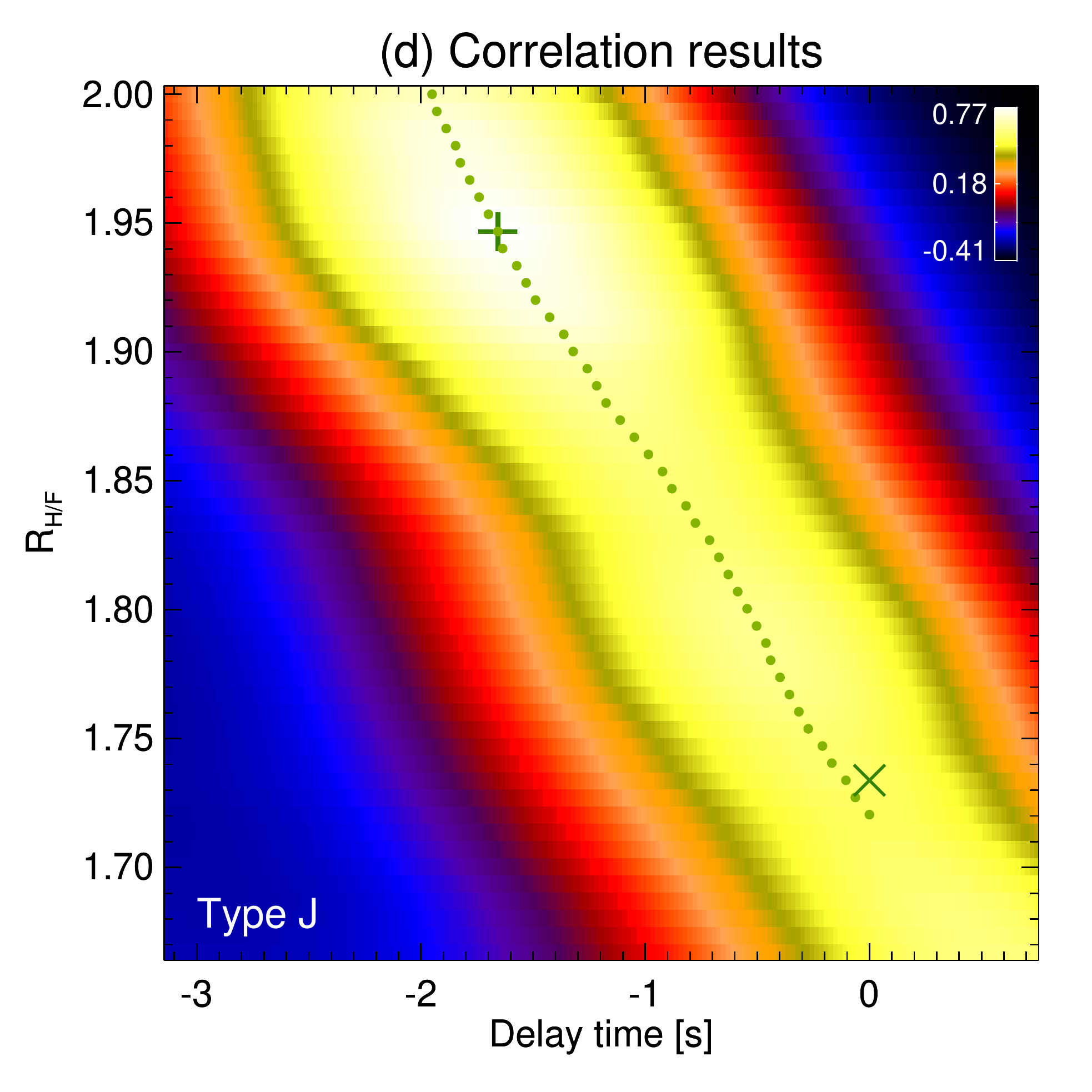}{0.41\textwidth}{ }}
\caption{Correlation results for type III and type J bursts: (a), (c) Enlarged dynamic spectra of the F (below) and H (upper) components. The solid rectangular box in the H spectrum will slip with given time and frequency lags. The selected F spectrum in the solid rectangular box is correlated with each slipped H spectrum. (b), (d) Two-dimensional cross correlation results between the selected F and each slipped H (spectral) boxes. The peak correlation coefficients at each frequency ratio are shown by green dots. The maximal correlation coefficient is marked by the plus symbol. If there is no time delay, the frequency ratio at the peak correlation coefficient is marked by the cross symbol.
\label{fig-cross-step}}
\end{figure*}

\subsection{Overview of the type III burst}
\label{sec4-1-1}

The type III burst presented in Figure \ref{fig-spec} is composed of two branches---the F branch between 30-65 MHz and H branch between 30-72 MHz. The background is subtracted by using quiet periods prior to the bursts, and only the burst properties are analyzed. Their frequency and time resolutions are 12.2 kHz and 52.4 ms, respectively.

The type III-IIIb burst with F-H pairs is well analyzed in several papers. \cite{2017NatCo...8.1515K} demonstrated that radio wave propagation effects dominated the observed spatial characterization of radio burst images. Using a model developed by \cite{2019ApJ...884..122K} to quantitatively study radio-wave propagation in anisotropic density fluctuations, \cite{2020ApJ...905...43C} found that anisotropic scattering simulations can reproduce the observed time profiles, centroid locations, and source sizes of the type IIIb radio burst. From analysis of the dynamic spectrum, \cite{2018SoPh..293..115S} provided statistically significant properties of individual striae, \cite{2018ApJ...856...73C} explained that the striae fine structures were caused by the background density fluctuations, and \cite{2018ApJ...861...33K} demonstrated that the striae frequency drift can be modulated by a propagating fast wave train.
In this study, we focus on the delay time and frequency ratio between the F and H components.

The F emission started at around 11:56:54 UT and ended at around 11:56:59 UT. Each distinct stria between 30-40 MHz contributes to a mean striae lifetime of about 1 s, with longer duration times at lower frequencies. The same time range and twice the frequency range for the H branch (60-70 MHz) with respect to the F branch (30-35 MHz) are selected for analysis.
The H/F frequency ratios are measured in two ways in our study: from their drift rates and the cross correlations between F and H spectra. The delay times are derived from the peak time intervals of the F and H flux profiles and cross correlations between F and H branches.

\subsection{Frequency drift rates}
\label{sec4-1-2}

The frequency drift rate $\frac{df}{dt}$ for the F emission at the local plasma frequency is $\frac{df_\mathrm{F}}{dt}=\frac{df_\mathrm{pe}}{dt}$. Since $f_\mathrm{pe}=\sqrt{e^2 n (r)/\pi m_e}$ is a function of background electron density, then it can be written as $\frac{df_\mathrm{F}}{dt}=\frac{df_\mathrm{pe}}{dn(r)}\frac{dn(r)}{dr}\frac{dr}{dt}=\frac{f_\mathrm{pe}}{2n}\frac{dn(r)}{dr}\frac{dr}{dt}$. 
For the H emission at double the plasma frequency, the frequency drift rate is $\frac{df_\mathrm{H}}{dt}=\frac{d2f_\mathrm{pe}}{dn(r)}\frac{dn(r)}{dr}\frac{dr}{dt}=\frac{2f_\mathrm{pe}}{2n}\frac{dn(r)}{dr}\frac{dr}{dt}$.
Therefore, the theoretical frequency drift rate ratio between H and F emissions is $D_\mathrm{H/F}=\frac{df_\mathrm{H}}{dt}/\frac{df_\mathrm{F}}{dt}=2$, which can be used for evidence of the harmonic branches.

From the dynamic spectrum, the time profiles at each frequency are fitted with a 1D Gaussian function. The peak times at the maximum fitted flux at each frequency are marked. They follow a linear function over small frequency ranges. They are then fitted using a linear function of $f=\frac{df}{dt}t+C$, seen from the two green lines in the dynamic spectra (Figure \ref{fig-spec}) for both the F and H components. 
The drift rates with 1-sigma errors are $\frac{df^1_\mathrm{F}}{dt}=-5.68\pm0.16$ MHz s$^{-1}$ and $\frac{df^1_\mathrm{H}}{dt}=-10.30\pm0.12$ MHz s$^{-1}$ for the F and H components respectively. The drift rate of the H component is around twice that of the F component, which can be further deduced as evidence of a F-H pair.

Defining the drift rates for the F and H components as $D_\mathrm{F}=\frac{df_\mathrm{F}}{dt}$ and $D_\mathrm{H}=\frac{df_\mathrm{H}}{dt}$, then considering the drift rate functions $f_\mathrm{F}=D_\mathrm{F}t+C_\mathrm{F}$ and $f_\mathrm{H}=D_\mathrm{H}t+C_\mathrm{H}$, one can derive a function of their H/F frequency ratio, $R^\mathrm{dr}_\mathrm{H/F}=\frac{f_\mathrm{H}}{f_\mathrm{F}}=\frac{D_\mathrm{H}}{D_\mathrm{F}}+(C_\mathrm{H}-\frac{D_\mathrm{H}}{D_\mathrm{F}}C_\mathrm{F})f^{-1}_\mathrm{F}$.
Their H/F frequency ratios are then calculated between the two green lines from Figure \ref{fig-spec}. The frequency ratios range from 1.65 to 1.67 in the F frequency range of 30-35 MHz.

\subsection{Cross correlations between the F and H branches}
\label{sec4-1-3}

The H component is distinguished from the F, seen from Figure \ref{fig-cross-step}(a).
We select part of the F component in a frequency range of 30-35 MHz by considering that the cut-off frequency of the H emission is around 70 MHz.

In order to derive a cross correlation map, the frequency range and time range need to be selected for the H component. The 2D cross correlation is then calculated between the selected F component (the white box in the lower panel from Figure \ref{fig-cross-step}(a)) and the slipped H component (the white box, as an example at one instance, in the upper panel from Figure \ref{fig-cross-step}(a)).

The frequency ratio is set up to range from 1.5 to 2.0 with a ratio lag of 0.01. A frequency ratio of 1.5 corresponds to a frequency range of 45.0-52.5 MHz for the H, and a frequency ratio of 2.0 will determine a H frequency range of 60-70 MHz.
The time delay of the H is set up to range from -1.84 s to 1.26 s, which is the time difference between the start time of the H and the start time of the F component at 11:56:55.8 UT. The time step is set to be 0.05 s. 
We keep the same time interval between the start and ending times for both F and H components.
The time range for the F is fixed from 0 s (11:56:55.8 UT) to 3.10 s (11:56:58.9 UT) and the time range for the H is slipped and changed with each delay. 
For example, while the time delay for the H is -1 s, the H time range is from -1 s (11:56:54.8 UT) to 2.10 s (11:56:57.9 UT).

In order to search for any time-delays and frequency ratios between the F and H spectra, we create two-dimensional cross-correlation functions (CCFs) as follows:
\[
\mathrm{CCF}=\frac{\sum\limits_{ij}(X_{ij}-\overline{X})\times\sum\limits_{ij}(Y_{ij}-\overline{Y})}{\sqrt{(\sum\limits_{ij}X_{ij}-\overline{X})^2\times(\sum\limits_{ij}Y_{ij}-\overline{Y})^2}}
\]
Here $X_{ij}$ and $Y_{ij}$ are two sets of spectra of the F and H components.
We loop over all delay times and frequency ratios and compute the overlap and correlation for each shift. The effective correlation coefficients range from 0 to 1, meaning no correlation and maximum correlation, respectively. 
The cross correlation map can be seen in Figure \ref{fig-cross-step}(b).

Uncertainties of the time and frequency ratio lags are determined using intensity randomization subset sampling by taking an observed dynamic spectrum and creating 50 variations where the observed intensity is varied by $I \pm \delta I$. 
The background flux level before the burst at each frequency is taken as the uncertainty on the flux, around 1 sfu, similar to \cite{2017NatCo...8.1515K}. 
$\delta I$ is randomly taken from a normal distribution with a mean of zero, and a standard deviation of one. 
The average frequency ratio and delay time are obtained and the errors are taken from the sum of the standard deviations, delay time, frequency ratio resolutions and also the time and frequency resolutions of LOFAR.

\begin{figure*}[htbp!]
\gridline{\fig{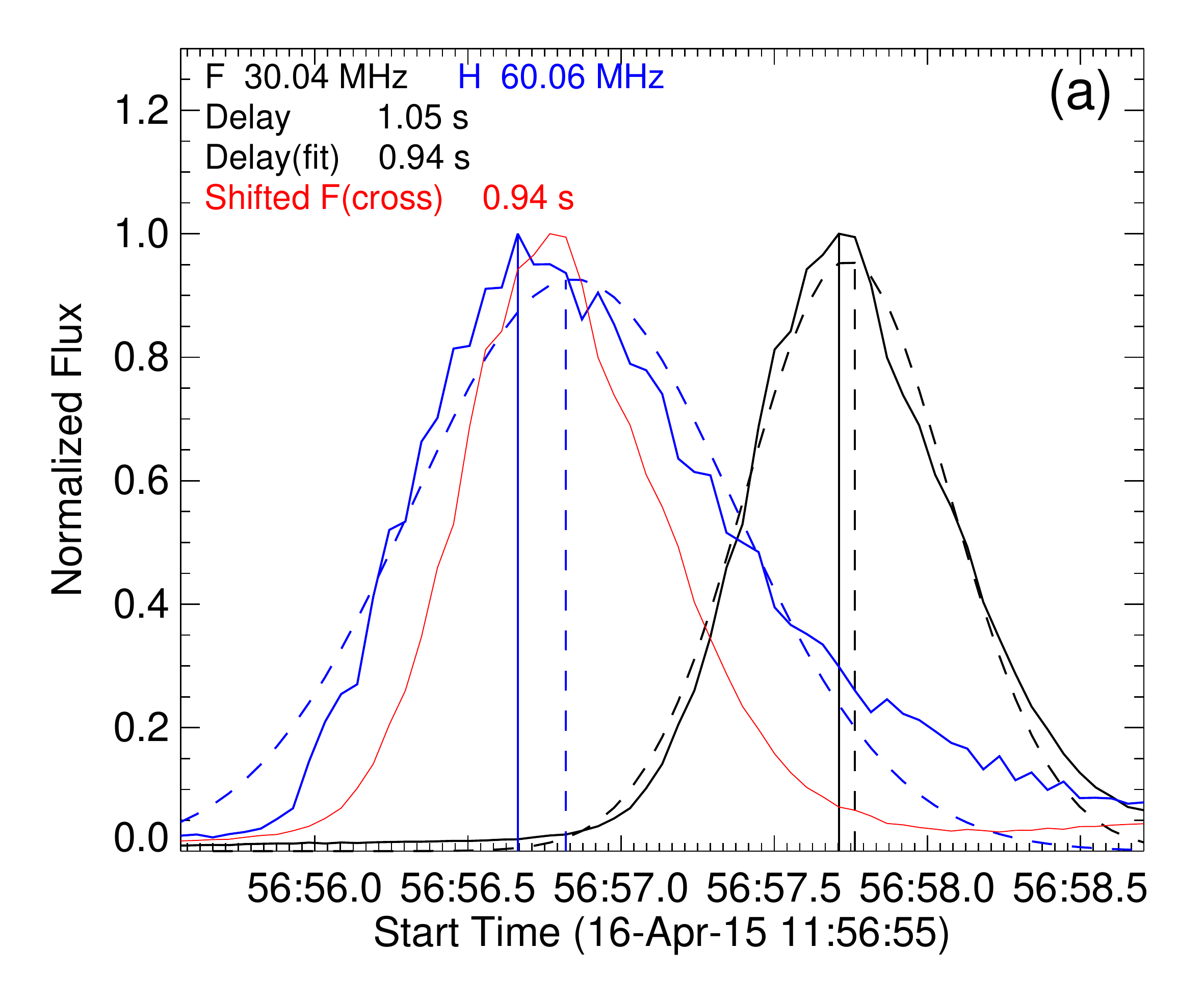}{0.4\textwidth}{ }
          \fig{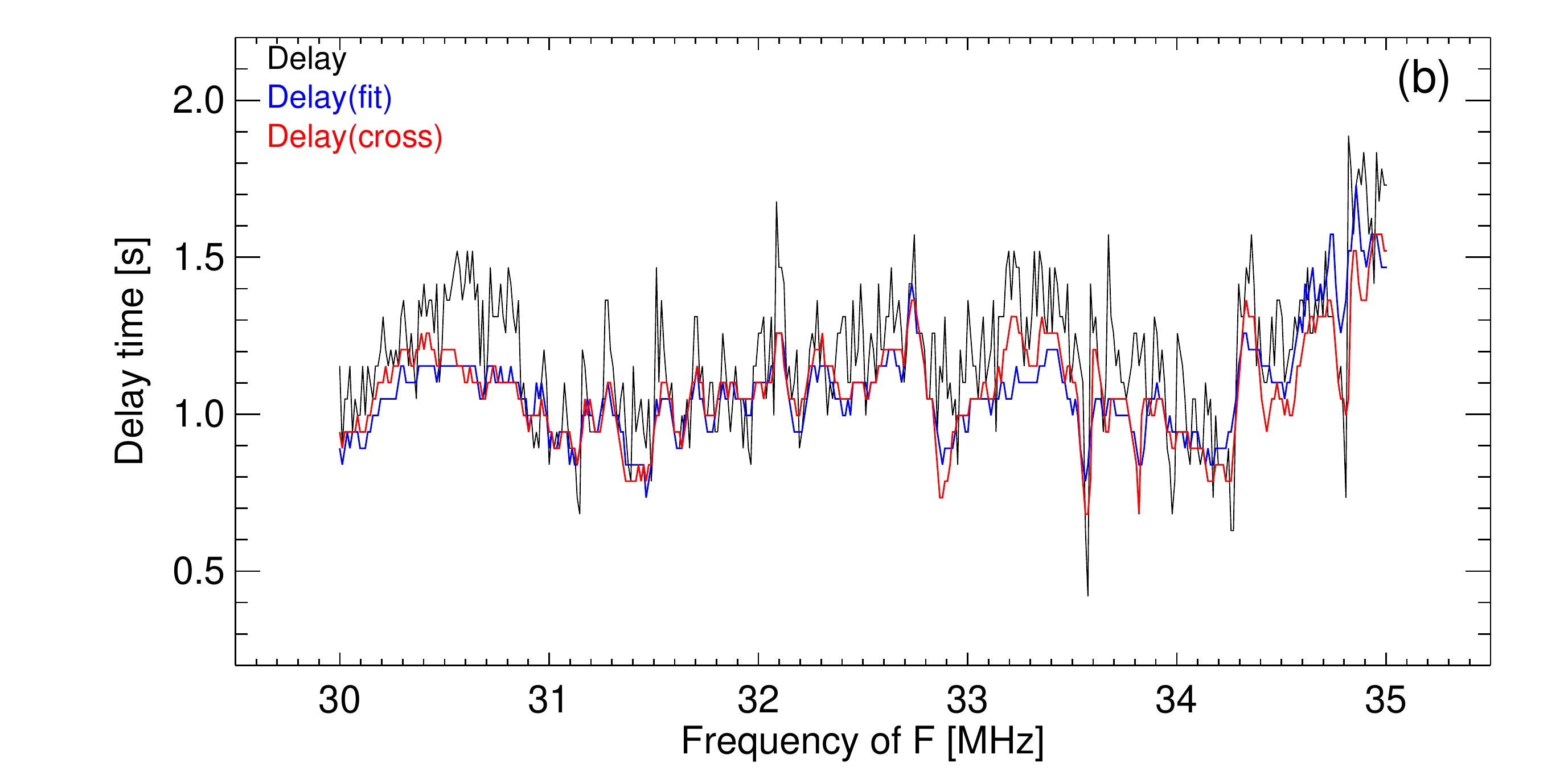}{0.65\textwidth}{ }}
\caption{Statistical delay times for the type III burst: (a) Normalized time profiles of the F and H emissions at 30 and 60 MHz, respectively. Their Gaussian fit profiles are shown by dashed lines. The peak times of observed and fitted time profiles are marked by the vertical solid and dashed lines, respectively. The time profile of the F component was shifted by a time lag of 0.94 s from the cross correlation between the F and H emissions (solid red line). (b) Statistical delay times from observations (black), fits (blue) and cross correlations (red) between F and H emissions between 30 MHz to 35 MHz.}
\label{fig-flux-delay}
\end{figure*}

From the cross correlation map, the peak correlation coefficients at each frequency ratio give the best-matched delay times (dot symbols).
The plus symbol shows the maximal correlation coefficient at a delay time of 1.00$\pm$ 0.05 s and a frequency ratio of 1.99$\pm$ 0.01 (seen from Figure \ref{fig-cross-step}(b)), which means that the H emission in the 60-70 MHz range is best matched with the F emissions in 30-35 MHz with a delay time of 1.00 s. In other words, the F emission is delayed by 1 s with respect to the H emission.
If there is no delay time between the F and H components, the frequency ratio is 1.66$\pm$ 0.01, as seen from the cross symbol in Figure \ref{fig-cross-step}(b).

\subsection{Statistical delay times}
\label{sec4-1-4}

The normalized time profiles at frequencies of 30 MHz (F, black solid line) and 60 MHz (H, blue solid line) and their Gaussian fitted profiles (dashed lines) are shown as an example in Figure \ref{fig-flux-delay}(a). The vertical lines mark the peak times of the flux curves.
The delay times derived from the intervals between the peak times of the original and fitted flux profiles between 30 MHz (F) and 60 MHz (H) are 1.05 s and 0.94 s, respectively. 
The time profiles of the F and H components are also cross correlated, which gives a delay time of 0.94 s for the maximum cross correlation coefficient. After correcting for a time lag of 0.94 s, the shifted F component is shown by the red solid line in Figure \ref{fig-flux-delay}(a). These three methods (estimations using the original, the fitted, and the cross correlation profiles) give similar delay times of around 1 second at 30 MHz.

We also statistically obtain the delay times from the flux profiles at each frequency between 30-35 MHz (F) and twice the frequency for the H components, shown in Figure \ref{fig-flux-delay}(b). It shows the averaged time delays for the original (black), fitted (blue), and the cross correlated (red) time profiles are 1.19 s, 1.14 s and 1.08 s, respectively.

\begin{table}
\centering
\begin{tabular}{c|clc}
\hline 
 & Type III Burst & Type J Burst\\
\hline 
$\overline{t}^\mathrm{F}_\mathrm{Dur}$ [s]& 1.27&  2.19\\
\hline 
$\overline{t}^\mathrm{H}_\mathrm{Dur}$ [s]& 1.19 &   2.10\\
\hline 
$f_\mathrm{mid}$ [MHz] &  32.5&  36.0\\
\hline 
$\frac{df_\mathrm{F}}{dt}$ [MHz/s]&  $-5.68\pm0.16$ &  $-2.38\pm0.11$\\
\hline 
$\frac{df_\mathrm{H}}{dt}$ [MHz/s] &  $-10.30\pm0.12$&   $-4.58\pm0.14$\\
\hline 
${R}^\mathrm{dr}_\mathrm{H/F}$ &  1.65-1.67 & 1.70-1.73\\
\hline 
$\overline{t}^\mathrm{pk}_\mathrm{d}$ [s] & 1.19&  2.04\\
\hline 
 $\overline{t}^\mathrm{fit}_\mathrm{d}$ [s] &  1.14& 1.97\\
\hline 
 $\overline{t}^\mathrm{corr}_\mathrm{d}$ [s] &  1.08& 1.91\\
\hline 
$R^\mathrm{corr(t_d=0)}_\mathrm{H/F}$ &  1.66&  1.73\\
\hline 
$R^\mathrm{corr(max)}_\mathrm{H/F}$& 1.99&  1.95\\
\hline 
$t^\mathrm{corr(max)}_\mathrm{d}$ [s]& 1.00& 1.67\\
\hline 
\end{tabular}
\caption{Characteristic parameters of the type III and type J bursts, including 
$\overline{t}^\mathrm{F}_\mathrm{Dur}$ (the averaged duration given by the FWHM from Gaussian fits for the F),  
$\overline{t}^\mathrm{H}_\mathrm{Dur}$ (averaged FWHM duration for the H), 
$f_\mathrm{mid}$ (middle frequency of the F component),
$\frac{df_\mathrm{F}}{dt}$ (drift rate of the F),
$\frac{df_\mathrm{H}}{dt}$ (drift rate of the H),
${R}^\mathrm{dr}_\mathrm{H/F}$ (frequency ratio calculated between the two drifting rate lines of the F and H), 
$\overline{t}^\mathrm{pk}_\mathrm{d}$ (averaged delay times calculated from the peak intervals of the flux curves of the F and H),
$\overline{t}^\mathrm{fit}_\mathrm{d}$ (averaged delay times derived from the peak intervals of the fitted flux curves),
$\overline{t}^\mathrm{corr}_\mathrm{d}$ (averaged delay times from the cross correlations between the F and H at each frequency),
$R^\mathrm{corr(t_d=0)}_\mathrm{H/F}$ (frequency ratio at the peak correlation coefficient for the case of no time delay),
$R^\mathrm{corr(max)}_\mathrm{H/F}$ (frequency ratio at the maximal correlation coefficient from the 2D cross correlation between the F and H spectra),
$t^\mathrm{corr(max)}_\mathrm{d}$ (delay time at the maximal correlation coefficient from the 2D cross correlation).}
\label{table1}
\end{table}

\subsection{Type J burst}
\label{sec4-1-6}

The F component of the type J burst was observed between 30-80 MHz, with the H component between 52-80 MHz. The F emission started at around 12:19:56 UT and ended with a long tail in the LOFAR observing frequency range. 

The same method to derive the frequency ratios and delay times are implemented for type J as for type III burst. All characteristic parameters are listed in Table \ref{table1}.
The frequency drift rate of the F branch is about $\frac{df^2_\mathrm{F}}{dt}=-2.38\pm0.11$ MHz s$^{-1}$ while that for the H branch is roughly twice that of the F branch at $\frac{df^2_\mathrm{H}}{dt}=-4.58\pm0.14$ MHzs$^{-1}$. 
Considering the lower drift rates than the normal type III burst and near "J" shape, the burst is identified as a variant of the type III burst called a type J burst \citep{2017A&A...606A.141R}.
The frequency ratios calculated from the drift rates range from 1.70 to 1.73 in the F frequency range of  33-39 MHz.

The results from cross correlations between the F and H components are shown in Figure \ref{fig-cross-step}(d).
It's worth noting that the lower cut-off frequency for the H is around 55 MHz, so the frequency ratio is set up to be from 1.67 (55 MHz/33 MHz) to 2.00. The delay times are set up from -3.15 s to 0.75 s. 
The frequency ratio and delay time are 1.95$\pm$ 0.01 s and 1.67$\pm$ 0.03 s at the maximal correlation coefficient from the two dimensional cross correlation map. The frequency ratio with no time delay is 1.73$\pm$ 0.01 at the peak correlation coefficient.

The delay times derived from peak time intervals of the original flux profiles, the fitted flux profiles, and the cross correlation between the F and H flux curves are 2.04 s, 1.97 s and 1.91 s, respectively, averaged  in a frequency range of 33-39 MHz (F) and twice those frequencies for the H components.


\begin{figure}[h!] 
\epsscale{1.1}
\plotone{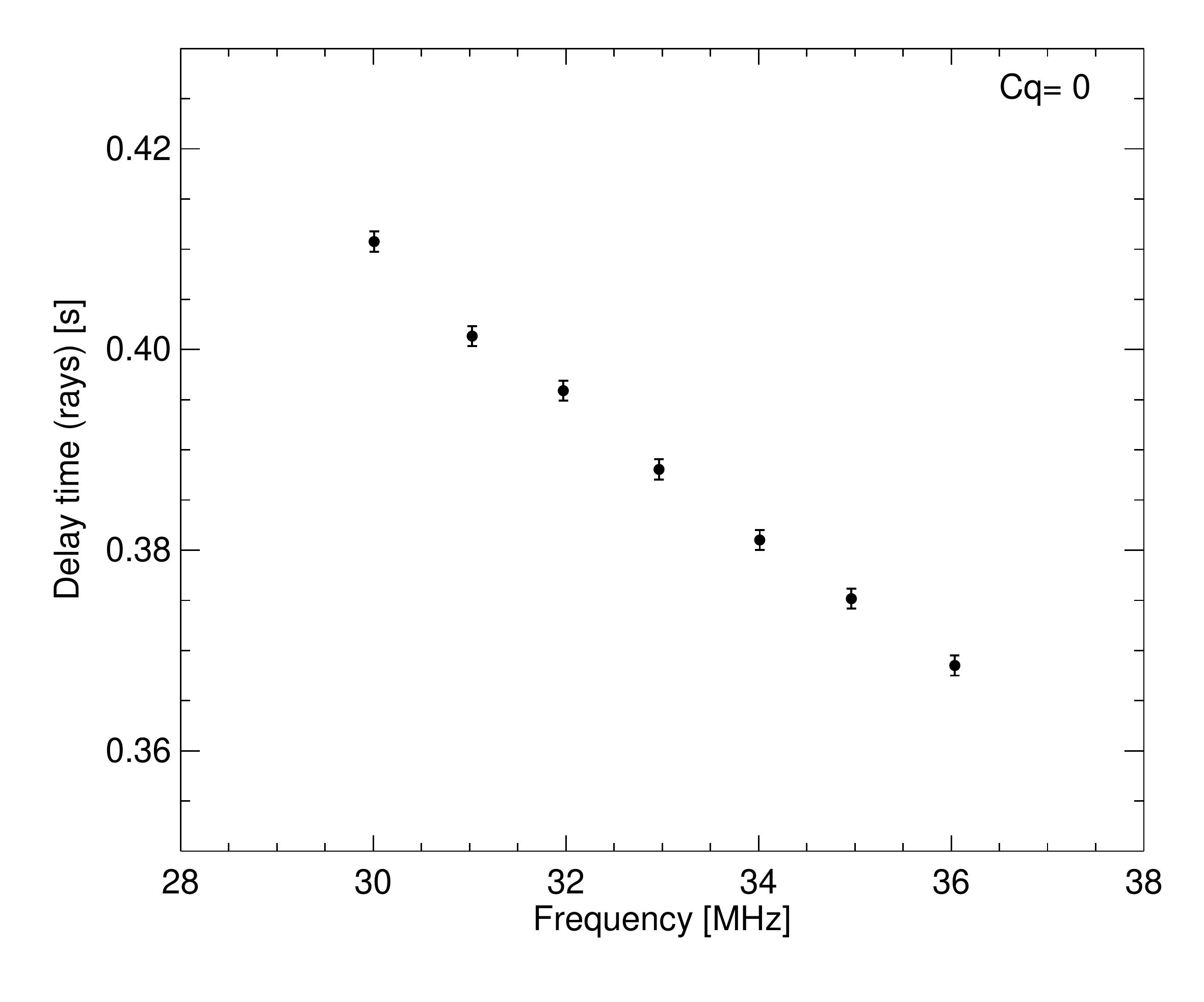}
\caption{Delay time as a function of frequency calculated using simulations without scattering: $C_q=$~0~$R_{\odot}^{-1}$, frequency ratio $R_\mathrm{H/F}$=1.82, and heliocentric angle $\theta=5^{\circ}$.}
\label{0sim_noscat}
\end{figure}

\begin{figure*}[htbp!]
\gridline{\fig{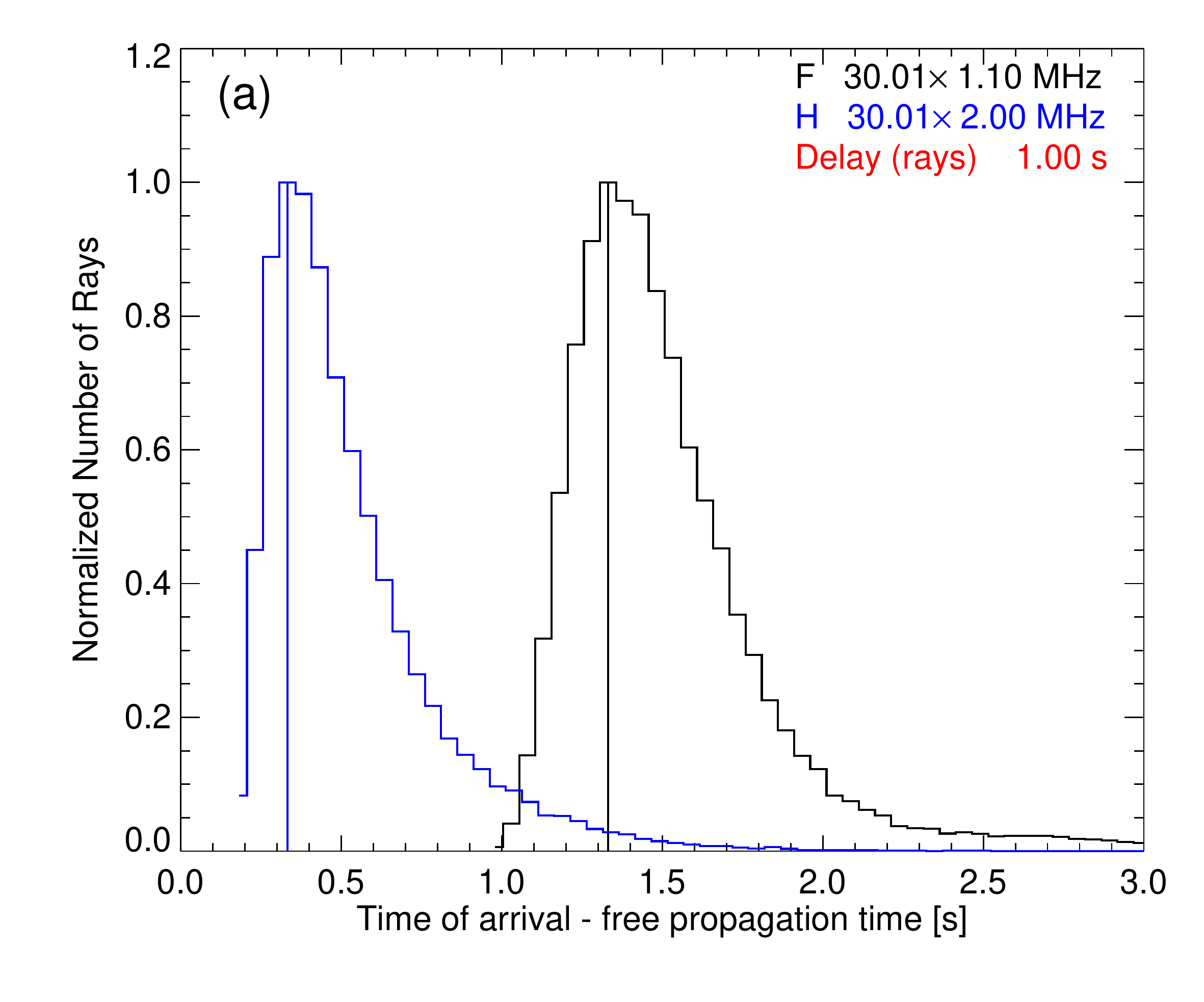}{0.45\textwidth}{ }
          \fig{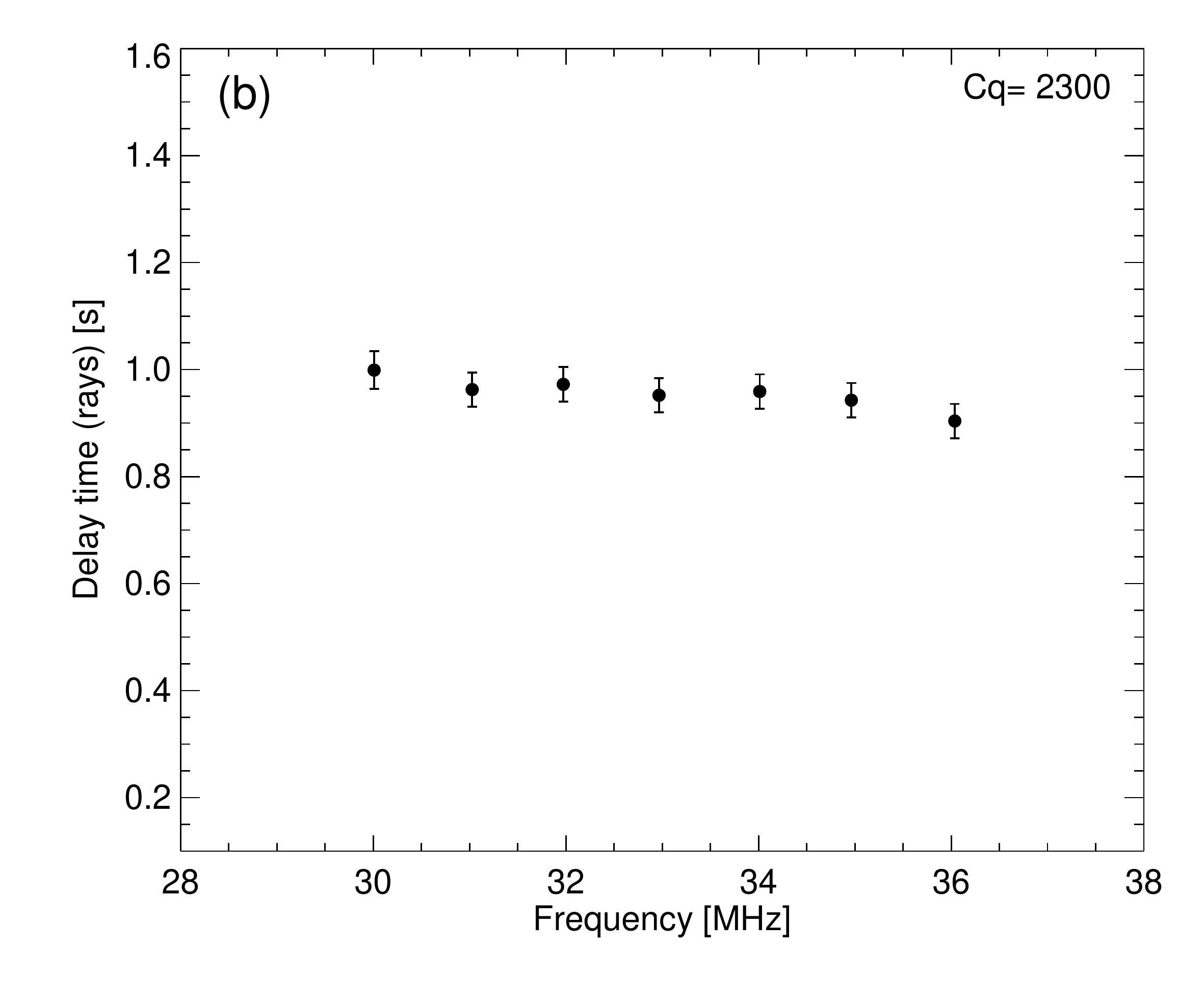}{0.45\textwidth}{ }}
\caption{Ray tracing simulation results: (a) Time profiles of both the F and H emissions at a local plasma frequency of 30 MHz, with a mean wavenumber of density fluctuations of $C_q=$~2300~$R_{\odot}^{-1}$, anisotropy $\alpha$=0.25, and heliocentric angle $\theta=5^{\circ}$. (b) Simulated delay times taken from time intervals between the peak intensities of the F and H emission in a frequency range from 30-36 MHz. The error bars represent the time bin width used for the histogram of the photon arrival times.}
\label{0sim_delay_step}
\end{figure*}

\begin{figure*}[htbp!]
\gridline{\fig{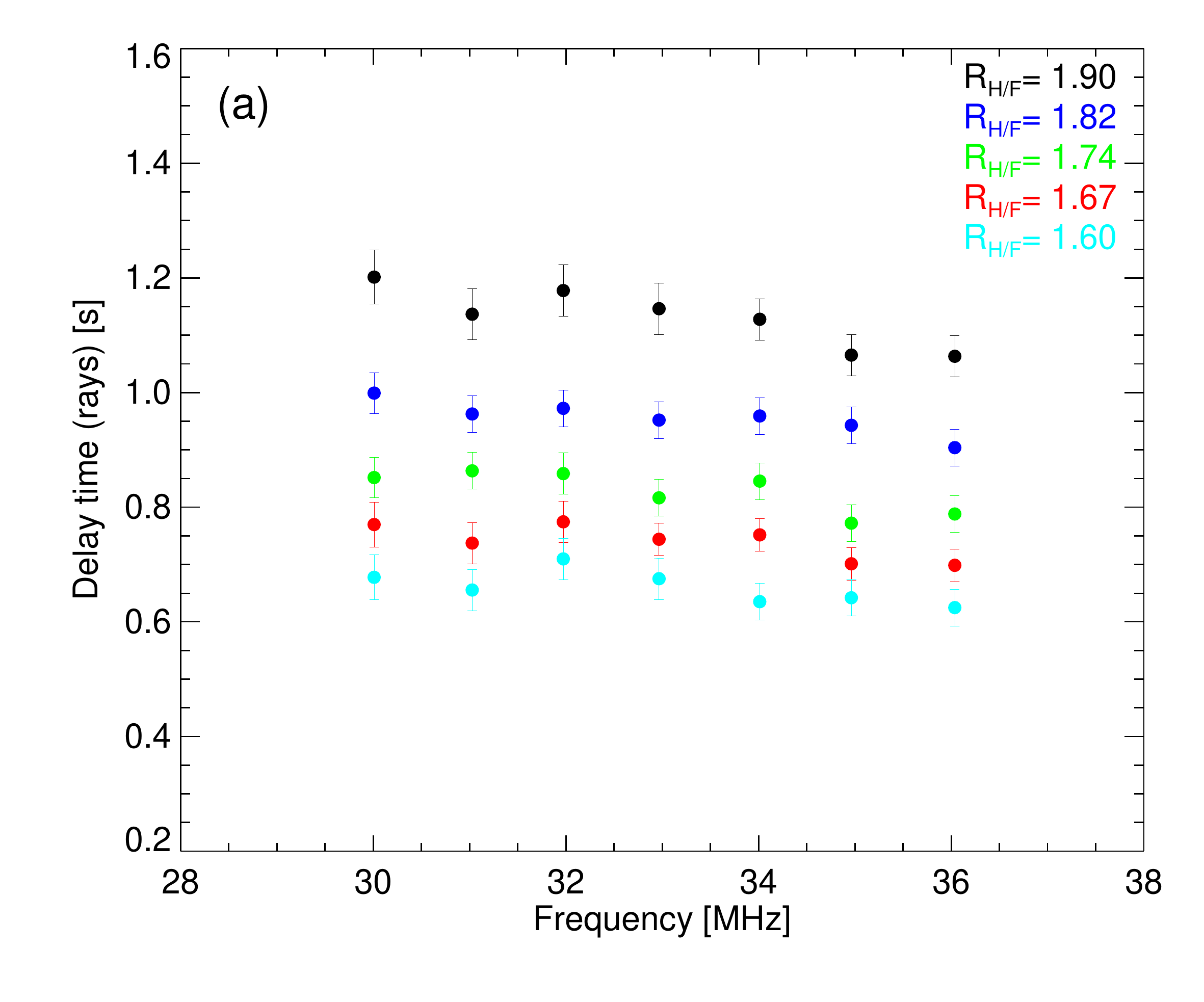}{0.45\textwidth}{ }
          \fig{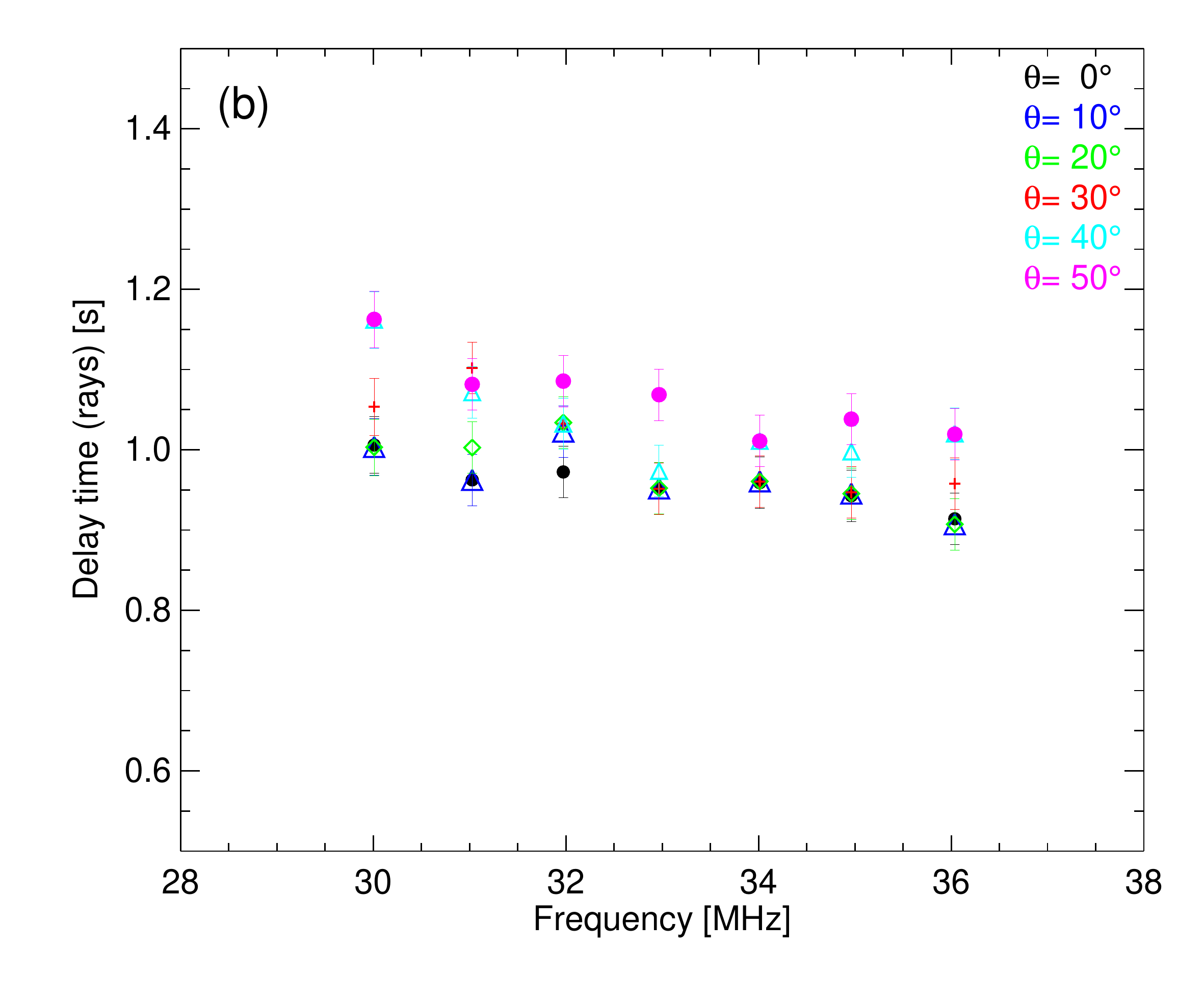}{0.45\textwidth}{ }}
\gridline{\fig{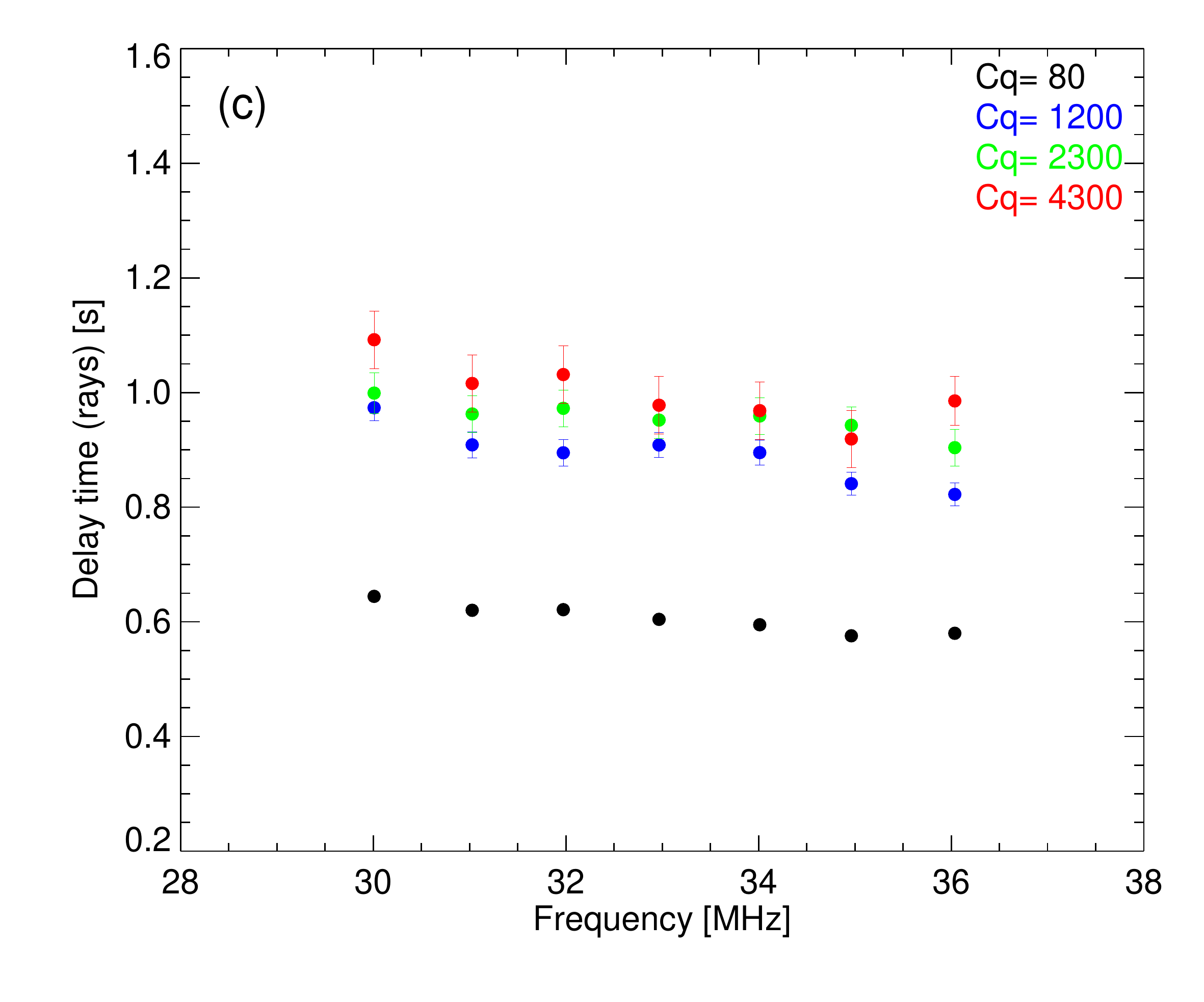}{0.45\textwidth}{ }
          \fig{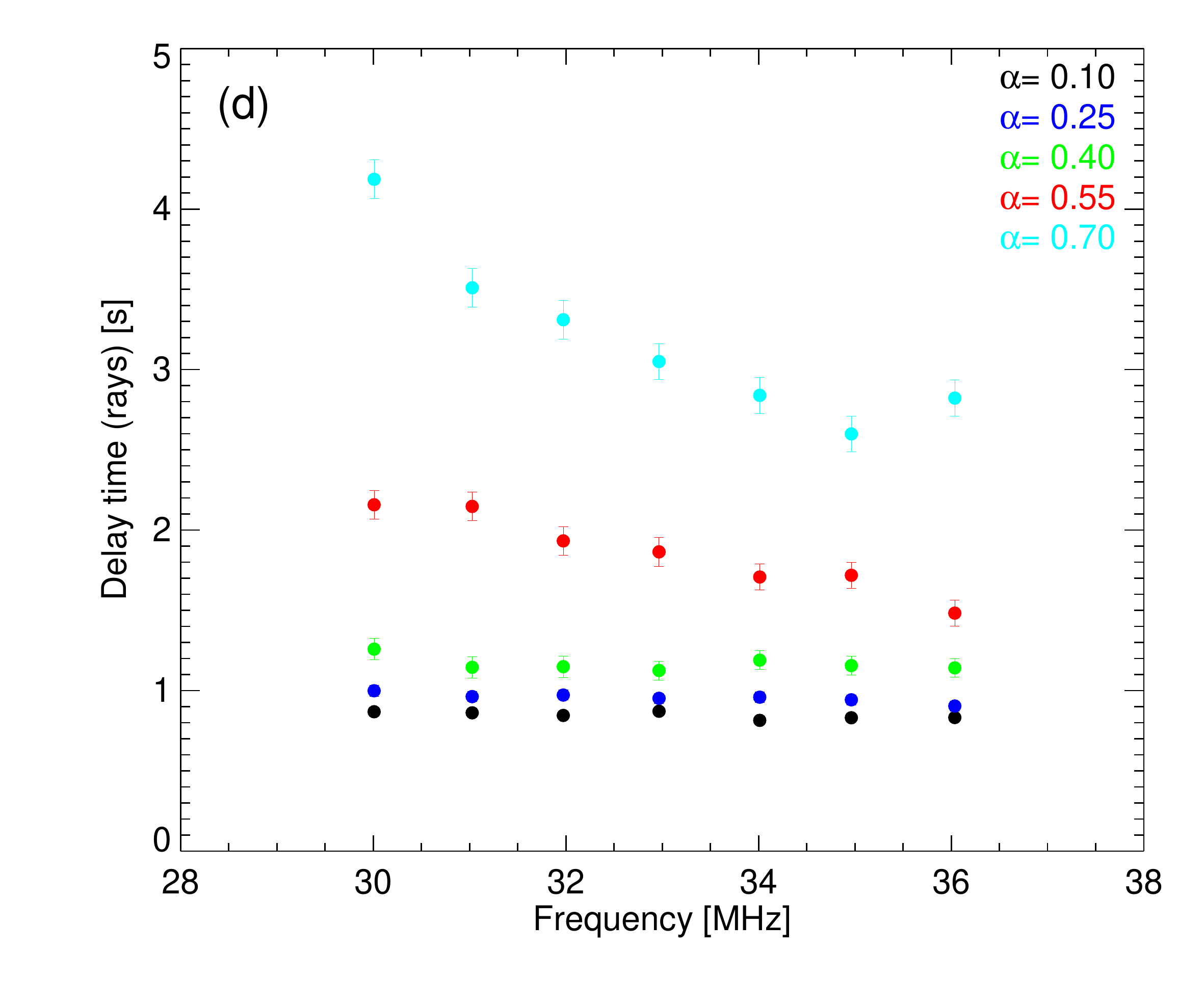}{0.45\textwidth}{ }}
\caption{Delay times from ray tracing simulation results for (a) multiple H/F frequency ratios ($R_\mathrm{H/F}$=1.60-1.90), (b) heliocentric angles ($\theta=0^{\circ}-50^{\circ}$), (c) multiple mean wavenumbers of the density fluctuations ($C_q$=80, 1200, 2300 and 4300 $R_{\odot}^{-1}$), and (d) anisotropy parameters ($\alpha$=0.10, 0.25, 0.40, 0.55 and 0.70). For the additional parameters assumed, (a) $\theta=5^{\circ}$, $C_q=$~2300~$R_{\odot}^{-1}$, $\alpha$=0.25, (b) $R_\mathrm{H/F}$=1.82, $C_q=$~2300~$R_{\odot}^{-1}$, $\alpha$=0.25, (c) $R_\mathrm{H/F}$=1.82, $\theta=5^{\circ}$, $\alpha$=0.25, (d) $R_\mathrm{H/F}$=1.82, $\theta=5^{\circ}$, $C_q=$~2300~$R_{\odot}^{-1}$, respectively. The error bars represent the time bin width for the simulated time profiles.}
\label{0sim_delay}
\end{figure*}

\begin{figure*}[htbp!]
\epsscale{1.1}
\plotone{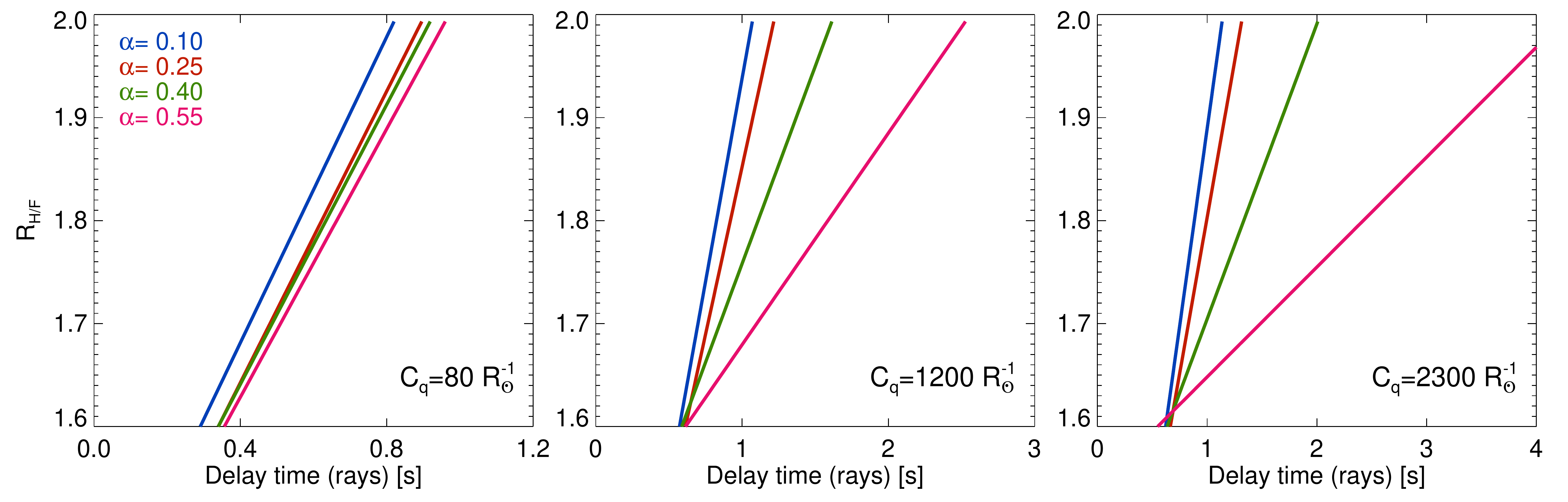}
\caption{Ray tracing simulation results at 30 MHz: frequency ratio versus delay time. The different colors in three panels represent a mixture of $C_q$=80, 1200, 2300 $R_{\odot}^{-1}$ and $\alpha$=0.10, 0.25, 0.40 and 0.55. 
\label{0Rfh_vs_delay_img}}
\end{figure*}

\section{Simulations of radio wave propagation}
\label{sec5}

The observed properties of radio waves, including time profiles, source positions, sizes, and emission directivity can be strongly affected by the inhomogeneous density fluctuations as they propagate through the turbulent corona.
There are very few studies on the time profiles that quantitatively investigate the delay times between F and H emissions resulting from their propagation through the turbulent coronal medium.
In order to quantitatively study the effects of radio wave propagation on the H/F frequency ratio and delay times between the F and H components, we use ray tracing simulations of radio wave propagation developed by \cite{2019ApJ...884..122K}.

\subsection{Brief introduction of the simulation set-up}
\label{sec5-1}

In the simulations, the radio waves are treated as a number of rays ($10^5$ in our case) with positions \textbf{r} and wavenumbers \textbf{k}. Initially, they are seen as a point source located at a given position which is related to the emitting frequency. Then the radio waves propagate in the turbulent corona and undergo refraction effects mainly caused by the large-scale density gradient, as well as scattering effects caused by small-scale density perturbations. Their positions and wave vectors are calculated from the numerical solutions of the Fokker-Planck equation and Hamilton's equations in an unmagnetized plasma (seen in \cite{2019ApJ...884..122K}).
All rays arrive at a sphere where the scattering is assumed to be negligible. Their arrival times, final positions and wave vectors are recorded to produce the time profiles and source images.

Importantly, the diffusion coefficient \citep[equation 14 in][]{2019ApJ...884..122K} is derived to describe the anisotropic scattering effects, which is related to the emitting frequency, the levels of the density fluctuation, the scale height of the turbulence, and the density anisotropy.
The simulated properties of the radio waves are mainly determined by the following: the frequency ratio over the local plasma frequency, the spectrum-weighted mean wavenumber of density fluctuations $C_q$ defined as $4\pi l_i^{-1/3}l_o^{-2/3}\epsilon^2=C_q r^{-0.88}$ (where $r$ is related to the emitting frequency in units of solar radii $R_\odot$, $\epsilon^2=\langle \delta n^2 \rangle/n^2$ is the variance of density fluctuations, and $l_i$ and $l_o$ give the inner and outer scales of the density turbulence), the anisotropic parameter $\alpha$, and the heliocentric angle $\theta$ between the line of sight and source position in the ecliptic plane.

\subsection{Cases without scattering}
\label{sec5-2}

Firstly, we investigate the delay times between F and H emissions without scattering effects (from small-scale density fluctuations) by assuming the $C_q$ number to be 0. 
From the dispersion relation for electromagnetic waves in an unmagnetized plasma, the group velocity $v_\mathrm{gr} = c^2k/\omega$ and ${\omega}^2={\omega}^2_\mathrm{pe}+k^2c^2$ will be different for the F emission at ${\omega}_\mathrm{pe}$ and H emission at 2${\omega}_\mathrm{pe}$. In this case, the time delay is caused by different group velocities of the F and H components and refraction effects from the large scale density perturbation. 

The F components emit at a frequency close to the local plasma frequency, which will be more strongly scattered than the H components that emit at a higher frequency. Thus, the peak time of the F will arrive later than that for the H. The simulated delay times are defined from the time interval between the peaks of the histograms showing the photon arrival times of the F and H emissions. 

In the case of no scattering on small-scale density fluctuations, the time bins for the F and H are 0.002 s so an error of $\sqrt{{(\delta t^\mathrm{F}}/2)^2+({\delta t^\mathrm{H}}/2)^2}\sim$ 0.001 s is directly considered. 
The statistical errors are from the finite number of photons in the simulations, such that a larger number of photons would give a smaller error in the simulated delay time.
The delay times vary from 0.41 $\pm$ 0.001 s to 0.37 $\pm$ 0.001 s at frequencies from 30 to 36 MHz, shown in Figure \ref{0sim_noscat}. 
It can be seen that the delay times caused without scattering (on small-scale density fluctuations) are apparently too small compared to the delay times from type III and type J burst observations. 
In the following, the delay times are investigated using multiple simulation parameters for anisotropic scattering which are necessary in order to explain solar radio emissions at meter to kilometer wavelengths, as shown by previous studies \citep{2019ApJ...884..122K, 2020ApJ...898...94K, 2020ApJ...905...43C, 2021A&A...656A..34M, 2021ApJ...909..195Z, 2021ApJ...917L..32C}.

\subsection{Delay times for multiple simulation parameters}
\label{sec5-3}

We take the same simulation parameters from \cite{2020ApJ...905...43C}, in which they found that the observed time profiles, source sizes, and motion of the type III-IIIb burst at 32 MHz (the same type III burst in  Figure \ref{fig-spec}) can be simultaneously explained with anisotropic radio-wave scattering due to turbulence with parameters $C_q=$~2300~$R_{\odot}^{-1}$, $\alpha$=0.25, at a heliocentric angle of $\theta=5^{\circ}$.

The simulated time profiles of both F and H components at a local plasma frequency of 30 MHz are presented by the histogram of the photon arrival times, shown in Figure \ref{0sim_delay_step}(a). 
We consider the F frequency 1.1$f_\mathrm{pe}$ and the H frequency 2$f_\mathrm{pe}$ (giving a frequency ratio of $R_\mathrm{H/F}$=1.82), the same as \cite{2020ApJ...905...43C}.
The F components undergo a stronger scattering effect and arrive later than the H components. As a result, the delay time is 1.00 $\pm$ 0.04 s, measured as the time between the peak times of the flux curves, which is close to the averaged delay time from the observation of the type III burst.

The simulated delay times in a frequency range from 30 to 36 MHz are presented in Figure \ref{0sim_delay_step}(b). The delay times and frequencies follow a linear relation from simulations, where the longer delay times correspond to emissions at lower frequencies. The delay times are roughly 1.00 $\pm$ 0.04 s at 30 MHz and 0.90 $\pm$ 0.03 s at 36 MHz.

The frequency ratios of the F component are simulated as being 1.05, 1.10, 1.15, 1.20 and 1.25 times the local plasma frequency $f_\mathrm{pe}$ while the frequency ratio of the H component is set to be 2$f_\mathrm{pe}$, which give H/F frequency ratios of 1.90, 1.82, 1.74, 1.67 and 1.60 respectively. The simulated delay times for multiple H/F frequency ratios are shown in Figure \ref{0sim_delay}(a). They have a linear relationship with frequency for all assumed H/F frequency ratios. The delay times are about 1.20 $\pm$ 0.05 s and 0.68 $\pm$ 0.04 s respectively for the H/F frequency ratios of 1.90 and 1.60 at $f_\mathrm{pe}=30$ MHz. The delay time is longer for a higher H/F frequency ratio, which is reasonable because radio emission emitting closer to the plasma frequency undergoes stronger scattering and thus incurs a longer delay.

The delay times for multiple heliocentric angles $\theta$ of the source varying from 0 to 50 degrees are represented by the different colors in Figure \ref{0sim_delay}(b). The delay times change from 1.01 $\pm$ 0.04 s ($0^{\circ}$) to 1.16 $\pm$ 0.04 s ($50^{\circ}$) at $f_\mathrm{pe}=30$ MHz for a H/F frequency ratio of 1.82. It seems that the heliocentric angle has a weak influence on the delay times which become only slightly extended for larger heliocentric angles.

We also investigate the delay times for multiple levels of density fluctuations by simulating for multiple spectrum-weighted mean wavenumbers of density fluctuations (the $C_q$ parameter), which are combinations of the density fluctuation level and inner and outer scales of the density fluctuations. 
Delay times for $C_q$=80, 1200, 2300, 4300 ~$R_{\odot}^{-1}$ with an anisotropy parameter of 0.25 are shown in Figure \ref{0sim_delay}(c). The delay times are 0.64 $\pm$ 0.01 s to 1.09 $\pm$ 0.05 s at $f_\mathrm{pe}=30$ MHz, for $C_q$= 80 and 4300 ~$R_{\odot}^{-1}$, respectively. As expected, stronger density fluctuations for a larger $C_q$ number will result in stronger scattering and thus longer delay times.

Anisotropic density fluctuations that are predominantly in the perpendicular direction to the magnetic field are required to describe the observed solar radio bursts \citep{2017NatCo...8.1515K, 2019ApJ...884..122K, 2020ApJ...898...94K, 2020ApJ...905...43C, 2021A&A...656A..34M}. When $\alpha<1$, radio wave propagation aligns (to a larger degree) with the radial direction, leading to a narrower time profile. When $\alpha=1$, it represents the case of isotropic density fluctuations. Anisotropy parameters of $\alpha$=0.10, 0.25, 0.40, 0.55, 0.70 are considered, as shown in Figure \ref{0sim_delay}(d). Anisotropic scattering has a significant effect on the time profiles. Strong anisotropy can highly reduce the duration of the radio emissions and delay times between the F and H. Weak anisotropic scattering with $\alpha=0.70$ produces a delay of 4.19 $\pm$ 0.12 s, whereas the delay is reduced to 0.87 $\pm$ 0.01 s for strong anisotropic scattering with $\alpha=0.10$ at $f_\mathrm{pe}=30$ MHz.

\subsection{Frequency ratio and delay time}
\label{sec5-4}

While inputting the density fluctuation parameters $C_q$ and their anisotropy $\alpha$ in our numerical models, we can determine the frequency ratio vs. delay time at each frequency. Figure \ref{0Rfh_vs_delay_img} shows our predictions of the relation between frequency ratios and delay times for $C_q=$ [80, 1200, 2300] and $\alpha$=[0.10, 0.25, 0.40, 0.55] at 30 MHz.
Consequently, we make it feasible to quantitatively predict the relation between the frequency ratio and delay time between the F and H emissions by deducing the density fluctuation properties from the spectral and imaging radio observations and then applying those parameters in ray tracing simulations.

\section{Summary}
\label{sec6}

We presented the H/F frequency ratios and delay times of the F component from observations of type III and type J bursts with striae and compared them to those resulting from radio wave propagation simulations.

The striae in F and H components are expected to correlate with each other.
Cross correlations between the spectra of F and H with striae are carried out to find their best matched counterparts and can give the frequency ratio and delay times simultaneously. The statistical delay times averaged at 1.1 s and 2.0 s with frequency ratios of 1.99 and 1.95 for the type III and type J burst, respectively. 
Without correcting for the delay times, the frequency ratios at the same time are 1.66 and 1.73---which are significantly different from the theoretical prediction of 2.

For the plasma emission mechanism, both the F and H emissions are generated at the same coronal location but normally the H emissions arrive earlier than the F emissions. The earlier arrival of the H emissions may be caused by the combination of two effects: the faster group velocity and weaker scattering effects than those on the F emissions. 
We estimate the delay times caused by the different group velocities and propagation effects through ray-tracing simulations \citep{2019ApJ...884..122K}. From our simulations, we quantitatively show the delay times between the F and H emissions for multiple H/F frequency ratios, heliocentric angles, density fluctuation levels, and anisotropy parameters. 
When there is no scattering from small-scale density fluctuations, the delay time is caused by the different group velocities and the refraction effects from large scale density fluctuations. Such delay time is estimated to be around 0.41 s at a local plasma frequency of 30 MHz, which is not sufficient to explain the observed delay time. If we adopt the same characteristic parameters as in the ray tracing simulations that successfully reproduced the observed properties of the same type IIIb burst analysed in \cite{2020ApJ...905...43C}, the simulated delay time ($\sim$ 1.00$\pm0.10$ s at 30 MHz) between the F and H components is very close to the observed delay times derived from the original ($\sim$1.05 s at 30 MHz), fitted ($\sim$0.94 s at 30 MHz) and cross correlated ($\sim$0.94 s at 30 MHz) time profiles of the type IIIb burst, implying that propagation effects have a main contribution to the delay times between F and H emissions.

It may be deduced from the simulations in Figure \ref{0sim_delay}, that a stronger scattering environment and weaker anisotropy give a longer delay time. From the context of observations, stronger scattering and weaker anisotropy would cause a longer burst duration, which itself may then imply longer delay times, as seen in Table \ref{table1}. The delay times vary from one event to another, likely due to radio-wave propagation effects which vary with the coronal properties from event to event.
The observed F component is delayed with respect to the observed H component at each time point, which may be one of the reasons that the source positions of F and H components do not coincide at the same time.
The delay of the F component with respect to the H component could contribute to the fact that the F and H sources do not coincide when imaged, alongside the effects of radio-wave propagation which cause a larger shift of the F sources away from their true position compared to the H sources \citep{2018ApJ...868...79C, 2019ApJ...884..122K, 2020ApJ...905...43C}.

Radio-wave propagation effects lead to a delay of the F with respect to their H counterparts, producing an observed frequency ratio lower than the theoretical ratio of $\sim2$. Radio burst F-H pair observations with fine structures can be used to derive this delay time and retrieve the theoretical frequency ratio between striae counterparts. The delay time is dependent on the anisotropic turbulent conditions that vary between events. We show that the same parameters that reproduce the decay times and source sizes can also predict the delay times in radio-wave scattering simulations, offering a quantitative solution to the long-standing question of why different frequency ratios are often observed.

\begin{acknowledgments}
XC and EPK are supported by STFC consolidated grant ST/T000422/1. 
DLC and EPK are thankful to Dstl for the funding through the UK-France PhD Scheme (contract DSTLX-1000106007).  NC thanks CNES for its financial support. We gratefully acknowledge the UK-France collaboration grant provided by the British Council Hubert Curien Alliance Programme that contributed to the completion of this work. The authors acknowledge the support by the international team grant (\href{http://www.issibern.ch/teams/lofar/}{http://www.issibern.ch/teams/lofar/}) from ISSI Bern, Switzerland. This paper is based (in part) on data obtained from facilities of the International LOFAR Telescope (ILT) under project code LC8\_027. LOFAR \citep{2013A&A...556A...2V} is the Low-Frequency Array designed and constructed by ASTRON. It has observing, data processing, and data storage facilities in several countries, that are owned by various parties (each with their own funding sources), and that are collectively operated by the ILT Foundation under a joint scientific policy. The ILT resources have benefited from the following recent major funding sources: CNRS-INSU, Observatoire de Paris and Universit\'{e} d'Orl\'{e}ans, France; BMBF, MIWF-NRW, MPG, Germany; Science Foundation Ireland (SFI), Department of Business, Enterprise and Innovation (DBEI), Ireland; NWO, The Netherlands; The Science and Technology Facilities Council, UK; Ministry of Science and Higher Education, Poland. XC thanks NSFC Grants 12003048.
\end{acknowledgments}

\bibliography{fh}{}
\bibliographystyle{aasjournal}
\end{document}